\documentclass{aa}  

\usepackage[normalem]{ulem}
\usepackage{txfonts}
\usepackage{graphicx}
\usepackage{subfigure}
\usepackage{adjustbox}
\usepackage{hyperref}
\hypersetup{
    colorlinks=true,
    linkcolor=blue,
    filecolor=blue,      
    urlcolor=blue,
    citecolor=blue,
    }

\begin{document}

   \title{Electromagnetic signatures from accreting massive black hole binaries in time domain photometric surveys}


   \author{F. Cocchiararo
          \inst{1,2}\fnmsep\thanks{E-mail: f.cocchiararo@campus.unimib.it} 
          A. Franchini\inst{1,2,3}\fnmsep\thanks{E-mail: alessia.franchini@unimib.it}
          A. Lupi \inst{1,2,4}
          A. Sesana \inst{1,2}
          }

   \institute{
    Dipartimento di Fisica "G. Occhialini", Università degli Studi di Milano-Bicocca, Piazza della Scienza 3, 20126 Milano, Italy             
    \and
    INFN, Sezione di Milano-Bicocca, Piazza della Scienza 3, 20126 Milano, Italy
    \and
    Universität Zürich, Institut für Astrophysik, Winterthurerstrasse 190, CH-8057 Zürich, Switzerland
    \and
    DiSAT, Università degli Studi dell’Insubria, via Valleggio 11, I-22100 Como, Italy\\
       }

   \date{}

 
\abstract
{We study spectral and time variability of accreting massive black hole binaries (MBHBs) at milli-parsec separations surrounded by a geometrically thin circumbinary disc. To this end, we present the first computation of the expected spectral energy distribution (SED) and light curves (LCs) from 3D hyper-Lagrangian resolution hydrodynamic simulations of these systems. We model binaries with total mass of $10^6$ M$_\odot$, eccentricities $e=0,\, 0.9$ and mass ratio $q=0.1,\, 1$. The circumbinary disc has initial aspect ratio of 0.1, features an adiabatic equation of state, and evolves under the effect of viscous heating, black body cooling and self gravity. To construct the SED, we consider black body emission from each element of the disc and we add a posteriori an X-ray corona with a luminosity proportional to that of the mini-discs that form around each individual black hole. We find significant variability of the SED, especially at high energies, which translates into LCs displaying distinctive modulations of a factor of $\approx 2$ in optical and of $\approx 10$ in UV and X-rays.
We analyse in detail the flux variability in the optical band which will be probed by the Vera Rubin Observatory (VRO). We find clear modulations on the orbital period and half of the orbital period in all systems.
Only in equal mass binaries, we find an additional, longer timescale modulation, associated to an over-density forming at the inner edge of the circumbinary disc (commonly referred to as lump).
When considering the VRO flux limit and nominal survey duration, we find that equal mass, circular binaries are unlikely to be identified, due to the lack of prominent peaks in their Fourier spectra. Conversely, unequal mass and/or eccentric binaries can be singled out up to $z\approx 0.5$ (for systems with $L_{\rm bol}\approx10^{42}$ erg s$^{-1}$) and $z\approx 2$ (for systems with $L_{\rm bol}\approx10^{44}$ erg s$^{-1}$). Identifying electromagnetic signatures of MBHBs at separations $\sim 10^{-4}-10^{-2}$ pc is of paramount importance to understand the physics of the gravitational wave (GW) sources of the future
Laser Interferometer Space Antenna, and to pin down the origin of the GW background (GWB) observed in pulsar timing arrays. }

   \keywords{Accretion, accretion disks --
                Hydrodynamics --
                quasars: supermassive black holes 
               }
   
\titlerunning{EM signatures from accreting MBHBs in time domain photometric surveys}
   \authorrunning{F. Cocchiararo et. al}
   \maketitle
%
\section{Introduction}

The existence of a massive black hole (MBH) at the centre of most, if not all, galaxies is a well established observational fact \citep[][and references therein]{Kormendy2013}. In the hierarchical clustering framework of structure formation, galaxies grow by accreting material from the intergalactic medium and by merging with other galaxies. Galaxy mergers are expected to result in the formation of MBH binaries (MBHBs). Indeed, during these merging events, the MBHs hosted at the centre of the parent galaxies migrate toward the centre of the merger remnant due to the dynamical friction mechanism \citep{Chandrasekhar1943}.
At parsec separation, where the two MBHs bind together to form a binary \citep{Begelman1980}, dynamical friction becomes inefficient in bringing further together the binary components \citep{Mayer2007b,Dotti2012}. 
Since gravitational waves (GWs) are efficient only at binary separations below $\sim 10^{-3} \rm \, pc$ , it is necessary to invoke one or more astrophysical mechanisms to shrink the binary further down, to the point where GWs can efficiently drive the merger. 
The most explored mechanisms involve the interaction of the binary with either its stellar \citep{Quinlan1996,Khan2011,2011ApJ...732L..26P} or gaseous \citep{ArmitageNatarajan2002,Escala2005,Lodato2009,Cuadra2009} environment. While the evolution driven by stars is always found to shrink the binary due to the energy and angular momentum exchange with stars in three-body stellar encounters \citep[e.g.][]{Sesana2007,Bortolas2021}, the effect of the interaction with the gaseous environment has recently been widely debated.
The effect of the circumbinary disc has been investigated extensively using 2D \citep{Munoz2019,Duffell2019,Tiede2020, Siwek2023} and 3D \citep{HeathNixon2020,Franchini2021,Franchini2022,Franchini2023} hydrodynamical simulations. The general consensus is that, although relatively warm discs with aspect ratio around 0.1 have been found to expand the binary, shrinking is usually promoted by cold less viscous circumbinary discs.

As a result of the interaction with the gaseous environment in which they reside, MBHBs are expected to produce distinctive observational electromagnetic (EM) signatures.
Currently known candidates at sub-parsec separations have been inferred from their EM emission \citep{Bogdanovic2022} through either photometric measurements of quasi-periodic variability both in optical \citep{Graham2015,Charisi2016, Liu2019, 2020MNRAS.499.2245C} and in $\gamma$-rays \citep[e.g.][]{2018A&A...615A.118S,2022arXiv221101894P} or spectroscopic measurements of offset broad emission lines \citep{Bogdanovic2009, Dotti2009,
BorosonLauer2009, TangGrindlay2009, Decarli2010, Barrows2011, Tsalmantza2011, Eracleous2012,Tsai2013}.
Convincing evidence that these sources are indeed MBHBs is still missing, both due to the lack of a firm theoretical understanding of the expected distinctive emission signatures of these systems and because the observed features can be accommodated by alternative explanations \citep[see, e.g.,][]{1994ApJS...90....1E,2016MNRAS.461.3145V}.

These observationally elusive MBHBs are also the main targets of current and future GW experiments. The future space-based Laser Interferometer Space Antenna  \citep[LISA,][]{LISA2023}, to be launched in 2035, will survey the milli-hertz GW window, where it is expected to observe the late inspiral and mergers of MBHBs with masses in the range $10^4-10^7$M$_\odot$ everywhere in the Universe. At even lower frequencies, in the nano-hertz regime, precision timing of an ensemble of millisecond pulsars within a so-called pulsar timing array \citep[PTA,][]{Foster1990} is sensitive to GWs emitted by MBHBs of billions of solar masses at milli-pc separations \citep[see e.g.][]{Sesana2008}. Although GWs might be eventually needed for a firm detection of MBHBs, the identification of convincing EM counterparts will help pinning down the properties of the emitting systems, thus realising the promises of low frequency multimessenger astronomy. The importance of understanding the EM emission of MBHBs has became even more compelling following the recent detection of a signal compatible with a gravitational wave background (GWB) reported by the European and Indian PTA \citep[EPTA+InPTA,][]{2023arXiv230616214A,2023arXiv230616224A,2023arXiv230616225A,2023arXiv230616226A,2023arXiv230616227A,Smarra2023}, NANOGrav  \citep{nanograv2023,2023ApJ...951L...8A,2023ApJ...951L...9A,2023ApJ...951L..10A,2023ApJ...951L..11A}, Parkes PTA \citep[PPTA,][]{ppta2023} and Chinese PTA \citep[CPTA][]{cpta2023}. Although the properties of the signal are compatible with a cosmic population of MBHBs, its origin cannot yet be observationally established from the GW data alone, and identification of EM counterparts might aid confirming the origin of this signal.

Attempts at characterising the EM signatures from MBHBs at very small separations (i.e. tens to a hundred gravitational radii) has been performed by several authors \cite[e.g.][]{Tanaka2012, Tang2018, Ascoli2018,2022ApJ...928..187C,MajorKrauth2023}. The same regime of separations has been recently investigated in a series of 2D numerical simulations, indicating that the lump periodicity coming from the streams and the cavity wall is more noticeable in circular binaries being a possible smoking gun signature of circularity \citep{Westernacher2022}. Moreover, \cite{Westernacher2023} proposes that EM signatures produced by MBHBs could be different from single MBH quasi-periodic sources as a consequence of the formation of eccentric mini-discs, prompted by the tidal force field generated by the binary. 
The characterisation of the EM signatures at intermediate scales, i.e. separations $\sim 10^{-4}-10^{-2}$ pc from full 3D hydrodynamical simulations that resolve the thermal evolution of the circumbinary disc, is currently missing and is of fundamental importance for aiding the identification of the origin of the GWB signal found with PTA experiments. Furthermore, the modelling of these signatures is of paramount importance for the identification of MBHB candidates in time-domain surveys, that can constitute the cosmic population of precursors of the merging binaries that will be detected by LISA.

In this work, we compute for the first time the spectral energy distributions (SEDs) and the multi-wavelength light curves (LCs) from 3D numerical simulations with hyper-Lagrangian refinement of milli-parsec scale binaries. We take into account the thermodynamic evolution of the gas using a  radiative cooling prescription in the form of black-body radiation. We also include the disc self-gravity, which is usually neglected, and the Shakura-Sunyaev prescription for viscosity \citep{ShakuraSunyaev1973}. 
We model binaries with eccentricity $e=0,\,0.9$ and mass ratio $q = 0.1,\,1$ and then compute their emitted spectra and light curves. 
We also add to the spectrum 
the contribution from a thermal corona, based on the assumption that its emission is proportional to the radiation emitted by the discs around each binary component (also called mini-discs). We place the simulated binary at different redshifts $z=\,0.1,\,0.4,\,0.7$, and analyse the observed flux in different frequency bands, mainly focusing on the optical band which will be probed by the upcoming Vera Rubin Observatory  \citep[VRO;][]{LSST2009}. As an exercise, we then assume the source to be two orders of magnitude brighter, but still preserving the same light-curve, and we investigate observability with the VRO at higher redshifts, $z=\,1,\,2,\,3$. 
Finally, we analyse the time evolution of the disc and binary properties such as the aspect ratio $H/R$ (where $H$ is the disc vertical height and $R$ is the distance from the central MBH in the disc-plane), the semi-major axis $a_0$ and the eccentricity $e$ of the binaries.

The paper is organised as follows. In Section \ref{NumericalPhysicalSetup} we describe the numerical details of the simulations and the physical parameters we use, the circumbinary disc physics assumptions and the thermal emission calculation. We show and discuss the main results we obtained including the time evolution of the main binary and disc properties in Section \ref{Results}. Finally, we draw our conclusions and discuss possible observational implications in Section \ref{sec:conclusions}.

\section{Physical and numerical setup}
\label{NumericalPhysicalSetup}

\subsection{Binary and circumbinary disc model}
\label{discmodel}

We perform 3D hyper-Lagrangian resolution hydrodynamics simulations of the evolution of  the thermal circumbinary disc around an accreting MBHB. We use the 3D meshless finite mass (MFM) version of the code \textsc{GIZMO} \citep{Hopkins2015} combined with the adaptive particle splitting technique \citep[see][for details]{Franchini2022} to increase the resolution inside the cavity carved by the binary. 
We ran four simulations combining two values of the binary mass ratio, $q = 0.1,\,1$, and eccentricity, $e=0,\,0.9$, each spanning over 1300 binary orbital periods.
The only exception is the circular binary with mass ratio $q=0.1$, which we evolve for just over 1060 orbital periods since the disc reached a steady state after fewer orbits in this case.
The MBHBs have, in code unit, a total mass of $M_{\rm B} = M_{\rm 1} + M_{\rm 2} = 1$ where $M_{1}$ and $M_{2}$ are the masses of the primary and secondary black hole respectively, and initial semi-major axis $a_{\rm 0}=1$. Each MBH is modelled as a sink particle with accretion radius $R_{\rm sink} = 0.05a$. As the binary orbit is allowed to evolve with time \citep{Franchini2023}, conservation of mass, linear and angular momentum are ensured during each accretion event, in the same way it is done in the {\sc phantom} code \citep{bate1995}.
We initially model the circumbinary disc with $N=10^6$ gas particles for a total mass $M_{\rm D} = 0.01 M_{B}$. The mass is distributed with an initial surface density profile $\Sigma \propto R^{-1}$ and a radial extent between $R_{\rm in}=2a$ and $R_{\rm out}=10a$. The disc is coplanar with the binary orbit and has an initial aspect ratio $H/R = 0.1$ in all our simulations. We generated the disc initial conditions using the SPH code {\sc phantom} \citep{Price2017}.

In our model, the thermodynamic evolution of the gas follows an adiabatic equation of state with index $\gamma = 5/3$, so that the disc is allowed to heat and cool and we can capture the effect of shocks. 
In the Shakura-Sunyaev accretion disc model, the angular momentum transport within the disc is modelled with the $\alpha$ viscosity parameter which constitutes a simple parameterisation of the turbulence within the disc. This turbulence might be driven by the magneto rotational instability (MRI) in highly ionized discs \citep{balbushawley1991}. We therefore include viscosity using the Shakura-Sunyaev parameterisation with kinematic viscosity $\nu = \alpha c_{\rm s } H$, where $c_{\rm s}$ is the gas sound speed, and set $\alpha=0.1$ in all our simulations.
Since we do also include the disc self-gravity together with a cooling prescription, our discs may develop gravitational instabilities (GIs) that will eventually result in an additional source of angular momentum transport \citep{lodatorice2004,Lodatorice2005,lodato2007sg}.
To ensure the gravitational stability of the disc, we set the initial Toomre parameter $Q > 1$ \citep{Toomre1964}. 

The vast majority of existing numerical simulations that consider the disc self-gravity \citep{Gammie2001, Lodatorice2005,Cuadra2009,roedig2014,Franchini2021} make use of a simple cooling prescription that assumes gas to cool on a timescale $t_{\rm cool}$ that is a multiple $\beta_{\rm cool}$ of the dynamical time $t_{\rm dyn}=\Omega^{-1}$, i.e. $\beta_{\rm cool} = \Omega t_{\rm cool}$. 
In this work, we use a more realistic radiative cooling in the form of black-body radiation. The cooling rate is given by: 
\begin{equation}
        \label{eqn:cooling}
        \dot Q = \frac{8}{3} \frac{\sigma_{\rm SB} T_{\rm i}^4}{\kappa \Sigma}
\end{equation}
where $\sigma_{\rm SB}$ is the Stephan-Boltzmann constant, $\kappa$ is the opacity, $\Sigma$ is the disc surface density and $ T_{\rm i} $ is the temperature of each element.
We assume the opacity $\kappa$ to be a combination of the Kramer opacity $ \kappa_{\rm Kramer} \propto \rho T_{\rm i} ^{-7/2} $ and the electron-scattering opacity $ \kappa_{\rm es} = 0.2 (1+X) \, \mbox{g}\, \mbox{cm}^2$, with hydrogen mass fraction $ X=0.59$.
When scaled to physical units, all our simulations have a binary total mass $M_{\rm B}= 10^6 M_{\odot}$ and initial separation
$a=4.8 \cdot 10^{-4} \, \rm{pc} \simeq 1.2 \cdot 10^4 \, R_{\rm g}$, where $R_{\rm g} = GM_{\rm B}/c^2$ is the gravitational radius of the binary. Note that this implies that we can follow the gas only down to $R_{\rm sink}=0.05a \sim 600 R_{\rm g} \sim 100 R_{\rm ISCO}$. Therefore, we should bear in mind that in the following we are neglecting the emission in the region between $100 R_{\rm ISCO}$ and $R_{\rm ISCO}$ and the total luminosity of the system is likely going to be underestimated.
Choosing this particular semi-major axis value sets the orbital period of the binaries to 1 year. According to the observing time of a VRO survey of 10 years, our choice of semi-major axis implies that we can observe the flux emitted by the source over 10 orbital periods. 
In order to investigate the thermodynamics that governs our circumbinary discs, we measure the rate of change of the internal energy due to both heating (i.e. shocks, PdV work, and viscosity) and cooling processes, and we then compare it to the cooling rate in Eq. \ref{eqn:cooling}.
We define an effective $\beta_{\rm eff}$ as
\begin{equation}
  \label{eqn:betaeff}
  \beta_{\rm eff} = \frac{u}{\dot u} \Omega
\end{equation}
where $u$ is the particle internal energy. 
We find $\beta_{\rm eff}$ to depend on the radius and to reach a stable constant value $\beta_{\rm eff} \sim 5$ at $r=2a$ and $\beta_{\rm eff} \sim 1.7 \times 10^{3}$ at $r=10a$ after an initial transient phase of about $\sim 800 \, P_{\rm B}$ in all our simulations. We compare $\beta_{\rm eff}$ with the $\beta_{\rm cool}$ we measure from the black body cooling and we find that $\beta_{\rm cool} \ll \beta_{\rm eff}$. We can therefore conclude that the heating mechanisms ensure the disc stability against the black body cooling, which initially causes the transient phase.


\subsection{Emission model}
\label{Emission model}

For each gas particle $i$ we calculate the temperature $T_{\rm i }$ assuming that both gas and radiation pressure contribute to the hydrostatic equilibrium of the disc, thus numerically solving the following implicit  equation for $T_i$ at each resolution element
\begin{equation}
    \label{eqn:Ptot}
    P_{\rm tot} = P_{\rm gas} + P_{\rm rad} = \frac{\rho k_{\rm B} T_{\rm i}}{m_{\rm p}\mu} + \frac{4}{3}\frac{\sigma_{\rm SB} T_{\rm i}^4}{c}.
\end{equation}

We divide the disc temperature domain into a 2D matrix with dimensions $512 \times 512$ pixels in the $x$-$y$ plane, which coincides with the MBHB orbital plane. For each pixel, we compute the midplane temperature as the average temperature of all the particles within $z$ coordinate $-0.05 a < z < 0.05 a $, obtaining the midplane temperature matrix $T$. For a more detailed analysis of the mini-discs, we further divide the spatial domain from the sink radius of each MBH out to $r = 3 a$ into a 2D matrix of $512 \times 512$ pixels. For each pixel in this grid, we compute the midplane temperature $T$ using the same method outlined above. 
We then compute the effective temperature in optically thick approximation ($\kappa\Sigma > 1 $) in each element of both matrices as: 
\begin{equation}
    \label{eqn:Teff}
    T_{\rm eff}^4 = \frac{4}{3}\frac{T^4}{\kappa \Sigma}
\end{equation}
where $\Sigma$ is the surface density of each element of the matrix.

We obtained the flux emitted by each pixel using Planck's formula:
\begin{equation}
    \label{eqn:planck}
    dL_{\nu} \equiv B_{\nu }  \, {\rm d}\nu {\rm d}A  = \frac {2h \nu^3}{c^2}\, \frac{{\rm d}\nu}{  \exp\left({\frac{h\nu}{k_{\rm B}T_{\rm eff}}}\right) -1 } {\rm d}A 
\end{equation}
where $h$ is the Planck constant, $k_{\rm B}$ is the Boltzmann constant and d$A$ is the area of each element.
In order to analyse the different contributions to the emission from each part of the disc, we have divided the simulated domain into five different regions: the two mini-discs which extend from the sink radius of each component to the Roche Lobe size, the streams region that extends from outside the Roche Lobe out to $r = 3a$, an inner $3a < r < 5a $ and an outer $5a < r < 10a $ part of the disc. 
For each region, we compute the SED through the sum of each pixel flux obtained with Eq.~\eqref{eqn:planck}. 

We further add the thermal emission that is expected to originate from the corona using the correction to the bolometric luminosity in \cite{Duras2020} in the hard X-ray (i.e. $0.2 - 10 \, \rm{keV}$) band.
We assume that the corona emission is proportional to the total luminosity of the mini-discs through the correction factor $K_{\rm band}$, which in this case is defined as the ratio between the mini-discs luminosity $L_{\rm MDs}$ and the luminosity in a given spectral band $L_{\rm{band}}$, i.e. $K_{\rm band}= L_{\rm MDs}/L_{\rm band}$. 
Following \citep{Duras2020}, the X-ray correction factor $K_{\rm X}$ is given by:
\begin{equation}
    \label{eqn:kx}
    K_{\rm X}(L_{\rm MDs}) = a \left [ 1 + \left (\frac{\log(L_{\rm MDs}/L_{\odot})}{b} \right )^c \right ]
\end{equation}
where $a,b,c$ are best-fit parameters shown in Table 1 in \citep{Duras2020}.
Moreover, we assumed that the hot corona thermal emission produces a power-law spectrum with $\nu L_\nu\propto\nu^{-0.7}$ 
\citep{Regan2019}.

Since we ultimately want to obtain LCs to compare with future observations, we place the source at different distances from the observer and we compute the observed flux assuming isotropic emission as:

\begin{equation}
    \label{eqn:flux}
    F=\frac{L_{\rm band}}{4\pi d_{\rm L }},
\end{equation}
where $L_{\rm band}$ is the luminosity integrated over a specific range of frequencies and $d_{\rm L }$ is the luminosity distance \citep{Hogg2000}. 
We then analyse the changes in the observed flux over time in order to identify modulations that might signal the presence of a MBHB. 

In order to understand whether these changes in the observed flux are detectable, we need to simulate a realistic detection scenario, thus taking into account the sensitivity limit of the employed survey and its observation timespan. In particular, in this study we will focus on MBHB observability with the VRO. To infer observability of the systems we adopt a similar method to the one used by \cite{Kelly2019}.
We add to the simulated fluxes a Gaussian noise with variance given by the telescope $5 \sigma$ sensitivity in the considered band. We then perform an FFT onto a limited number of cycles (commensurate to a survey timespan of $\approx 10$yrs) and qualitatively identify whether prominent peaks in the spectrum appear and what their origin is. We defer a more thorough statistical study of the detectability of these features to a follow-up study.
The VRO $5 \sigma$ sensitivities in different bands are listed in Table \ref{tab:LSSTbands}. 

\begin{table}
\caption{Frequency bandsand 5-$\sigma$ flux sensitities of the VRO telescope's optical filters that we employ in the analysis of the emitted flux (\url{https://pstn-054.lsst.io/PSTN-054.pdf} for magnitude values details). }
    \resizebox{\columnwidth}{!}{%
	\begin{tabular}{lccccccr}
		\hline
        \noalign{\smallskip}
		 & Y & Z & I & R & G & U &  \\
        \noalign{\smallskip}
		\hline
        \noalign{\smallskip}
		lower & 2.83 & 3.25 & 3.67 & 4.34 & 5.43 & 7.5 & $\times 10^{14} \, \rm{ Hz }$ \\
        \noalign{\smallskip}
        upper & 3.16 & 3.67 & 4.34 & 5.43 & 7.5 & 8.57 & $\times 10^{14} \, \rm{ Hz }$ \\
        \noalign{\smallskip}
        $5\sigma$ sensitivity flux & 40.58 & 9.23 & 4.47 & 3.11 & 2.32 & 6.04 & $\times 10^{-15} \, \rm{ erg/s/cm^2 }$ \\
        \noalign{\smallskip}
        \hline
        \noalign{\smallskip}
        \noalign{\smallskip}
        \noalign{\smallskip}
	\end{tabular}}
    
    \label{tab:LSSTbands}

      
\end{table}

\begin{figure}
	\begin{center}
	\includegraphics[width=\columnwidth]{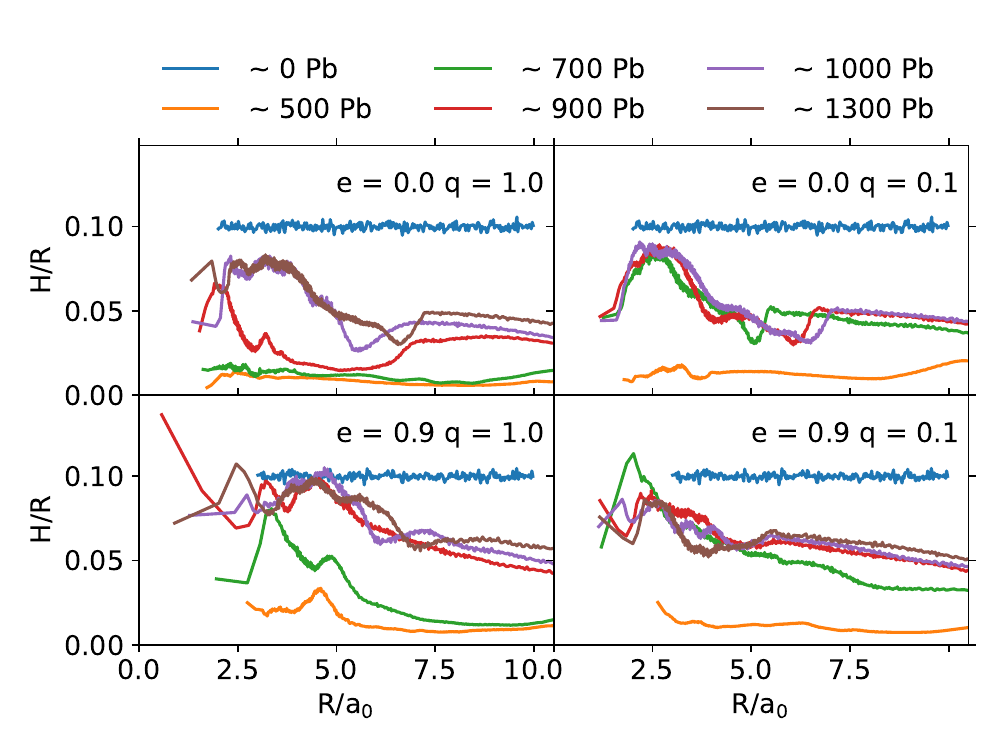}
    \caption{Time evolution of the disc aspect ratio H/R as a function of radius for circular binaries (top panel) and eccentric binaries (bottom panel) with mass ratio $q=1$ (left column) and $q=0.1$ (right column). For all the simulations, we report the thickness profile at different times with different colours. All the simulations start with $H/R=0.1$. The aspect ratio decreases during a transient phase and increase again reaching a constant value, similar for all the simulations: in the inner part of the disc $H/R \sim 0.08-0.09$ while in the outer regions of the disc $H/R \sim 0.04-0.06$. Unequal mass binaries stabilise their disc temperature faster compared to equal mass binaries.
    \label{fig:HR}
    }
 	\end{center}
\end{figure}

\section{Results}
\label{Results}

We here present the results we obtain from our numerical simulations. In particular, we show the time evolution of the disc aspect ratio $H/R$ and  briefly discuss the evolution of the binary orbital eccentricity and semi-major axis.  We then focus on the disc emission and on the LCs in various frequency bands and on the effect that the explored parameters (i.e. mass ratio, eccentricity and redshift) have on the detectability of the EM emission. 

\subsection{Disc and binary evolution}
\label{Results_HR_ae}

\begin{figure*}
    \includegraphics[width=\textwidth]{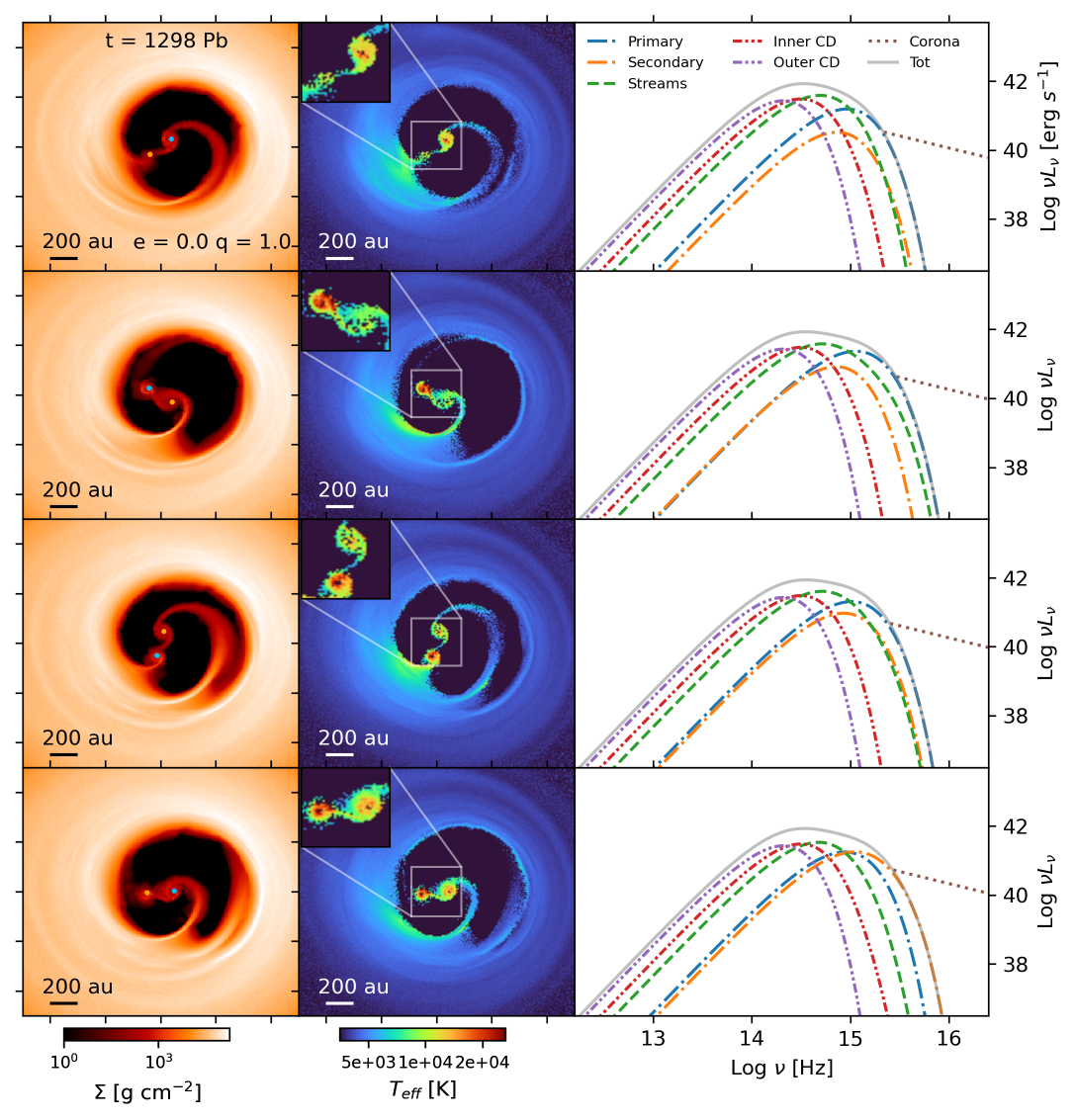} \\
    \caption{From the top to the bottom row: circular equal mass binary at four different orbital phases at time $t = 1298 \, P_{\rm b}$. The left, middle, and right panels in each row show the surface density map in the $x-y$ plane, the effective temperature map in the $x-y$ plane and the SED. The tick spacing on the $x$ and $y$ axes in the left and central panels is 2$a_0$, where $a_0$ is the initial binary semi-major axis.
    The surface density upper limit is set at $1.8 \times 10^5 \, \rm{g \, cm^{-2}}$. 
    In the SEDs, the contribution of the mini-discs around the primary and the secondary black hole is shown by the dot-dashed blue and orange lines respectively, the stream region is represented by the green dashed line while the inner and outer part of the circumbinary disc are represented by the red and purple dot-dashed lines respectively. The corona contribution is shown by the brown dotted line. The solid grey line shows the total emitted spectrum.  }
    \label{fig:SEDe0q1}
\end{figure*}
\begin{figure*}
    \includegraphics[width=\textwidth]{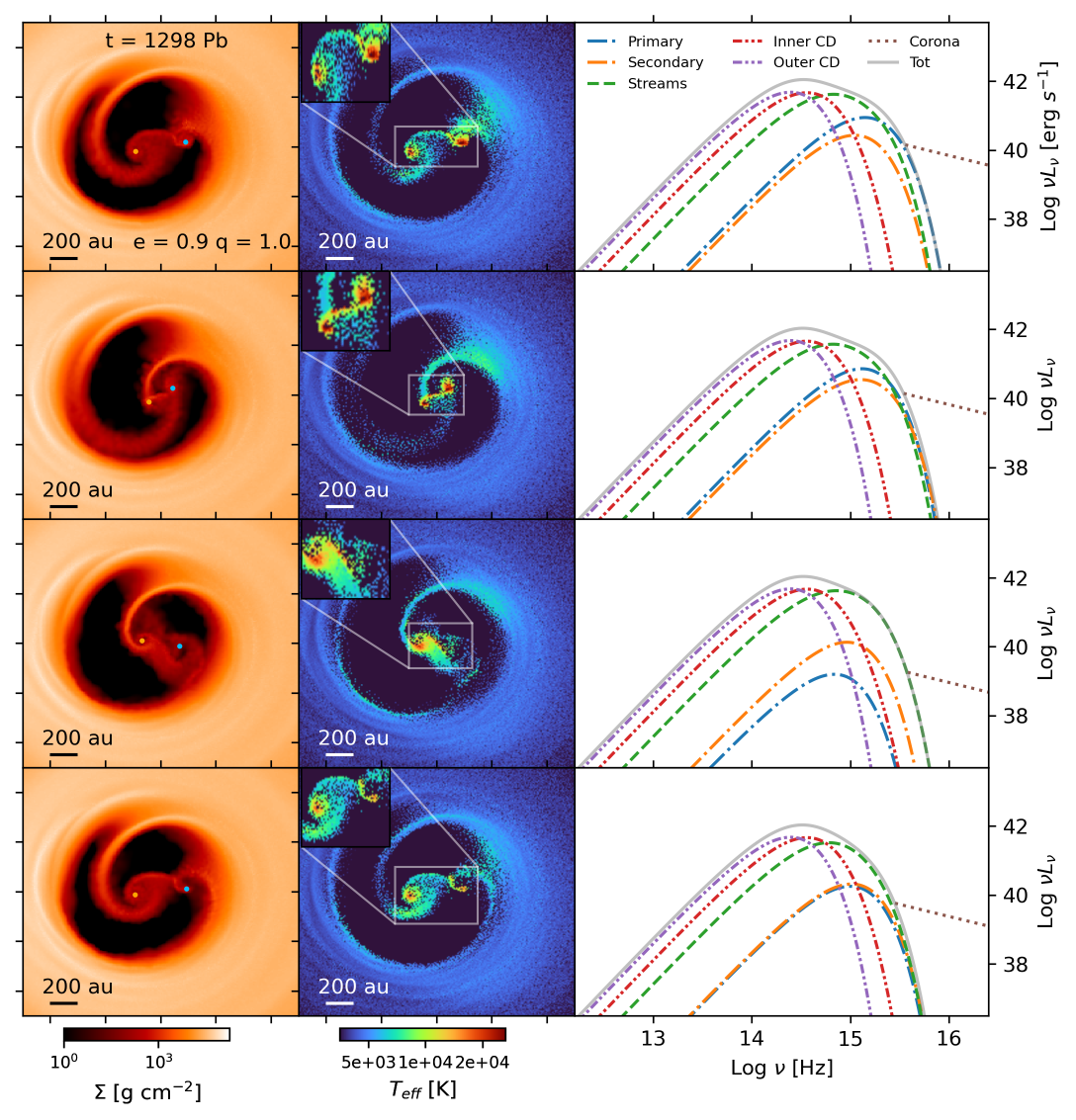} \\
    \caption{
    Same as Figure \ref{fig:SEDe0q1} but for the highly eccentric $e=0.9$ equal mass binary.}
    \label{fig:SEDe09q1} 
\end{figure*}
\begin{figure*}
    \includegraphics[width=\textwidth]{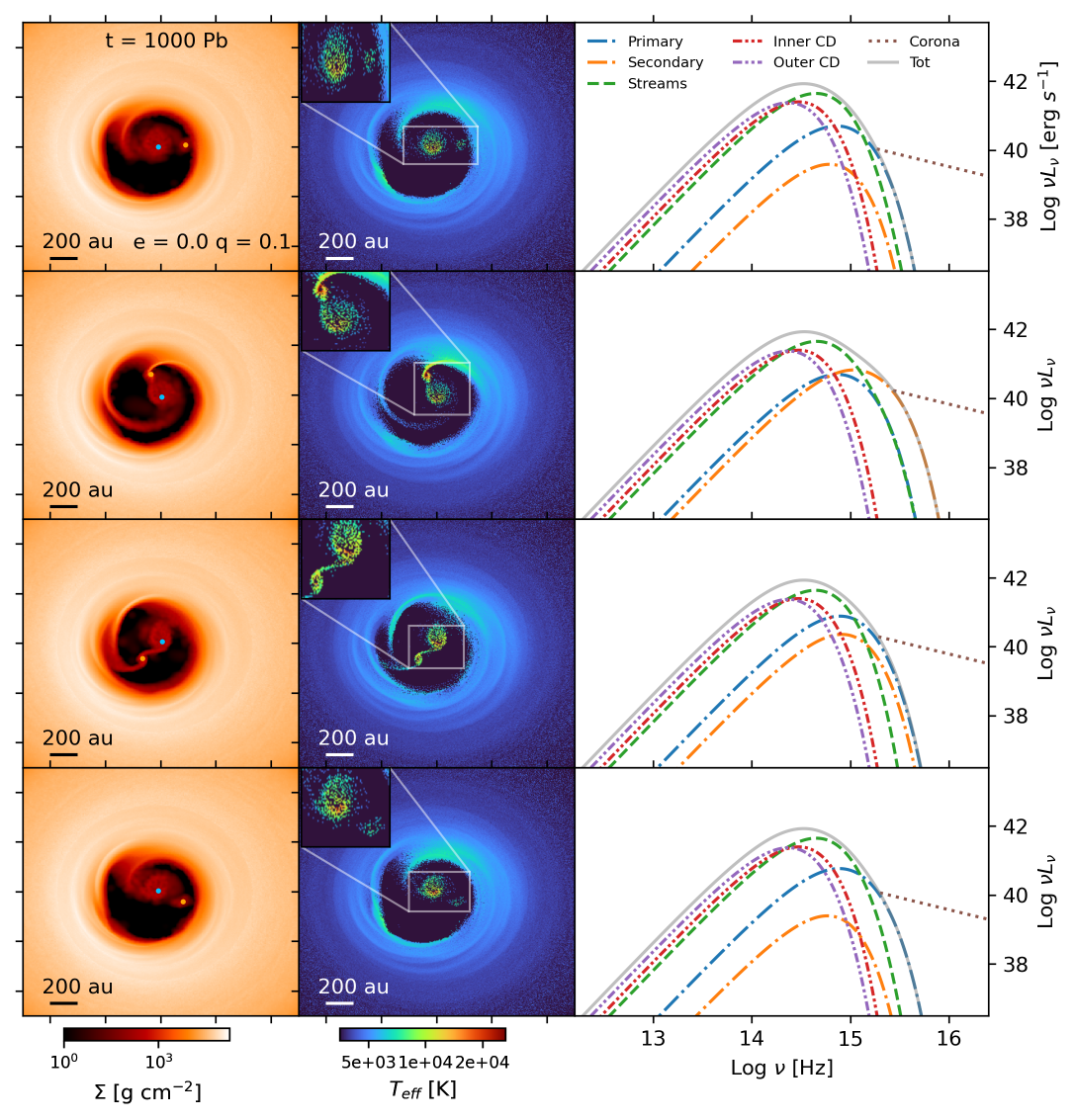} \\
    \caption{ Same as \ref{fig:SEDe0q1} but for a binary with  mass ratio $q=0.1$ and eccentricity $e=0$ and at time $t = 1000 \, P_{\rm b}$.}
    \label{fig:SEDe0q01}
\end{figure*}
\begin{figure*}
    \begin{center}
    \includegraphics[width=\textwidth]{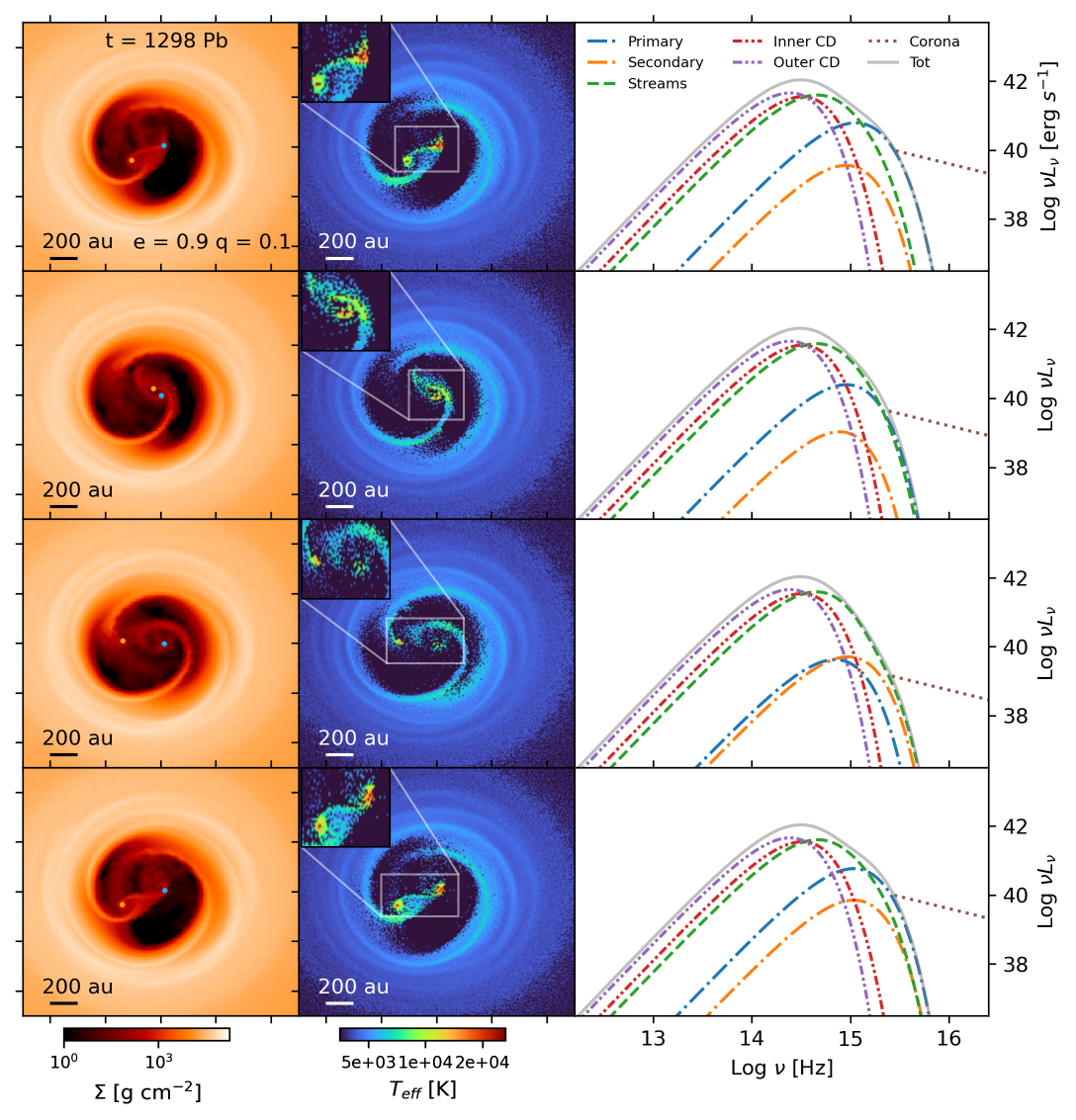} \\
    \caption{ Same as \ref{fig:SEDe0q1} but for a binary with  mass ratio $q=0.1$ and eccentricity $e=0.9$ and at time $t = 1298 \, P_{\rm b}$.}
    \label{fig:SEDe09q01}
    \end{center}
\end{figure*}
Since our numerical simulations are 3D and the gas is allowed to change its temperature with time, we can resolve gas shocks and investigate the effect they have on the final temperature of the disc.
As showed by previous simulations in the literature \citep{artymowicz1996, Hayasaki2007, Macfadyen2008,roedig2011, Shi2012, farris2014, Franchini2022, Westernacher2022, Westernacher2023}, part of the streams of gas that enter the cavity is flown back to the disc, creating shocks at the inner edge of the cavity wall that affect the disc aspect ratio. 
Figure \ref{fig:HR} shows the evolution of the aspect ratio profile at $t = 0, 500, 700, 900, 1000, 1300 \, P_{\rm B}$ for the circular cases (upper panels) and eccentric cases (lower panels) with mass ratio $q=1$ (left panels) and $q=0.1$ (right panels).
Since $Q > 1$, the disc cools down until $Q \sim 1$. Indeed, the aspect ratio decreases to $H/R \sim M_{\rm D }/M_{\rm B} \sim 0.01$ within the first few orbits.
In all our simulations, the disc then re-expands in the vertical direction and the aspect ratio increases, eventually reaching a stable profile. 
This is consistent with the disc temperature increase from an initial value at the inner edge of $T_{\rm eff}\sim 10^3 \, K$ to a final value of $T_{\rm eff}\sim 10^4 \, K$. 
Since gas particles are continuously flung back into the cavity wall, the shocks that occur in this region maintain the disc aspect ratio to $H/R\sim 0.08$, regardless of the choice of binary mass ratio and eccentricity. The radiative cooling in the outer parts of the disc is more efficient and the aspect ratio reaches eventually a value $H/R\sim 0.04$.
Equal mass binaries have a bigger quadrupole mass moment that exerts larger torques on the inner edge of the disc cavity, eventually producing more prominent streams and stronger shocks. As a consequence, discs around equal mass binaries reach a stable configuration after $\sim 1000 \, \rm{P_{B}}$ ($e=0$) and $\sim 900 \,\rm{P_{B}}$  ($e=0.9$), while it takes only  $\sim 700 \, \rm{P_{B}}$ for discs around unequal mass binaries, regardless of eccentricity.

The evolution of the binary semi-major axis depends on the contributions to the net torque from two distinct components: the gravitational torque exerted by the circumbinary disc on each binary component and the gas accretion torque \citep[see][for the detailed computation of the two contributions]{roedig2011}. 
Early numerical simulations of circumbinary discs showed the binary inspiral to be aided by the presence of the disc \citep{artymowicz1994,artymowicz1996,ArmitageNatarajan2002}.
The findings of more recent 2D, fixed binary orbit, hydrodynamic simulations show that the angular momentum transfer onto the binary is positive and this leads the expansion of the binary \citep{Munoz2019, Duffell2019, Munoz2020, Moody2019}. Using a similar numerical scheme, \cite{Tiede2020} found that the fate of the binary depends on the disc temperature, i.e. on the disc aspect ratio. They find the threshold for expansion to be $H/R>0.04$, while 3D smoothed particle hydrodynamics (SPH) simulations of locally isothermal discs find the threshold to be much higher, i.e. $H/R=0.2$ \cite{HeathNixon2020}. 
In a recent study, \cite{Franchini2022} showed, employing 3D MFM simulations where hyper-Lagrangian resolution was achieved via adaptive particle splitting (the same employed here), that the binary inspiral/outspiral depends also on the disc viscosity.
Numerical simulations that study the regime where the disc self-gravity can not be neglected, have found that the interaction between the binary and its gaseous disc leads the binary to shrink as a consequence of the time evolution of the disc temperature, regardless of its initial value \citep{Cuadra2009, roedig2012, Franchini2021}.

The interaction between the binary and its circumbinary disc changes the the binary eccentricity as well \citep{goldreich1980}. Based on analytial arguments, \cite{Artymowicz1991} shown that an equilibrium eccentricity should exist, which was later confirmed by the SPH simulations of \cite{roedig2011}, who derived an equilibrium eccentricity $0.5<e<0.8$ for comparable mass binaries, linking the precise value to the disc cavity size.
More recent 2D, fixed orbit, hydrodynamic simulations found equal mass binaries to reach an equilibrium eccentricity value around $\sim 0.45$ \citep{Zrake2021,DOrazio2021}. Using a very similar numerical scheme, \cite{Siwek2023} finds that binaries with mass ratios $q>0.2$ evolve towards an equilibrium eccentricity of $e\sim 0.5$.

While these latter works assumed the disc to be locally isothermal, we here allow the disc temperature to change with time due to PdV work heating and radiative cooling. 
We find the interaction between the binary and the circumbinary disc to cause the binary to shrink, regardless of the initial conditions. 
In particular, $a$ decreases by $ \sim 1\%$ over $\sim 1000$ orbits in all cases, except for the binary with $q=0.1$ and $e=0.9$. In this case, the binary initially experiences an expansion of $0.05\%$ over the first 600 orbits, likely because of the very high eccentricity that brings the lower mass MBH very close to the initial cavity edge. This is only a transient phase, as the binary then carves a larger cavity compared to the circular case and starts to shrink, transferring angular momentum to the circumbinary disc.
We find the eccentricity value to remain relatively constant during the whole simulation for initially eccentric binaries, while circular ones feature a mild eccentricity increase, reaching $e \sim 0.04$ in the equal mass case and $e \sim 0.06$ in the unequal mass case.

\begin{figure*}
    \centering
    \includegraphics[width=\columnwidth]{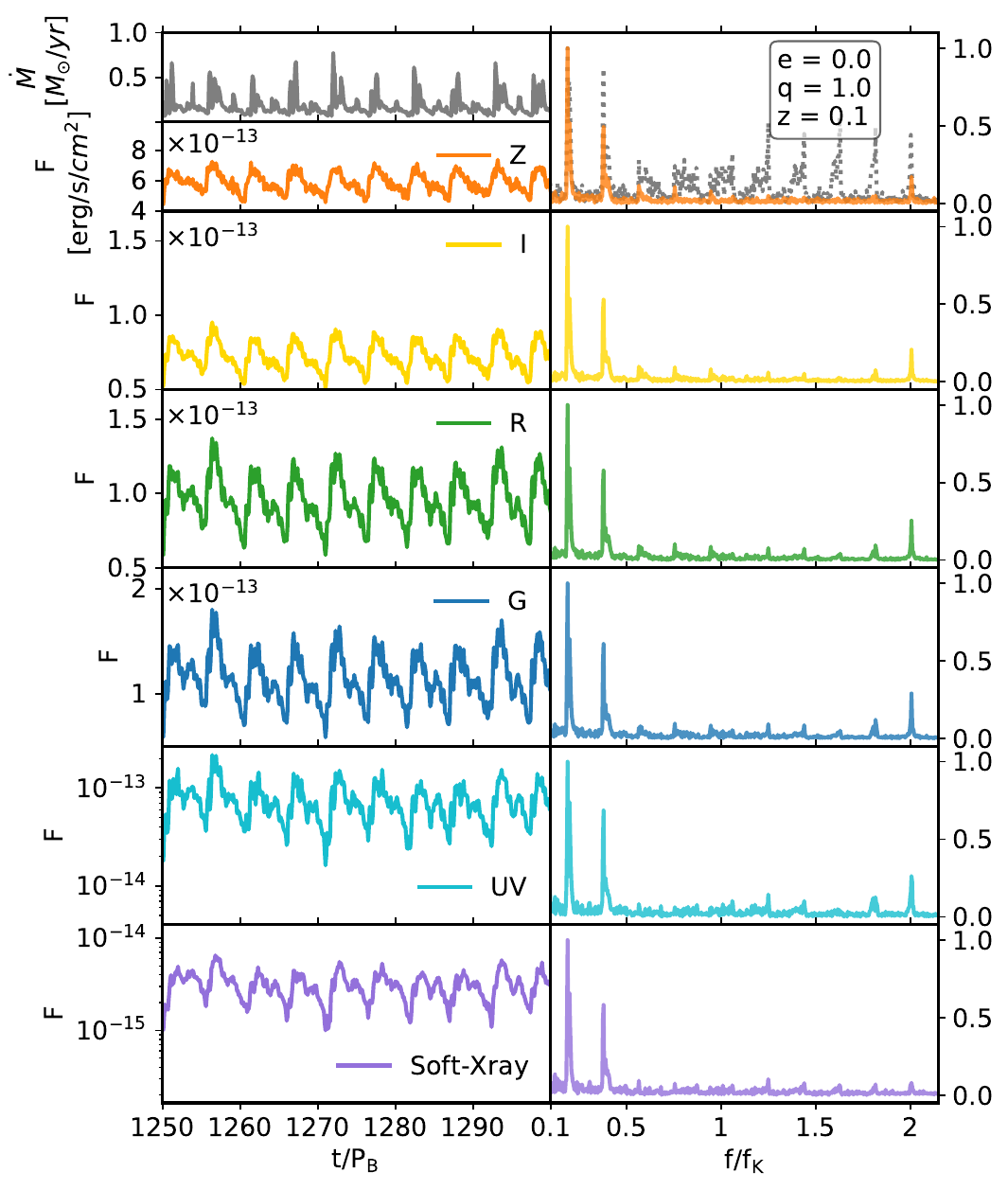}
    \includegraphics[width=\columnwidth]{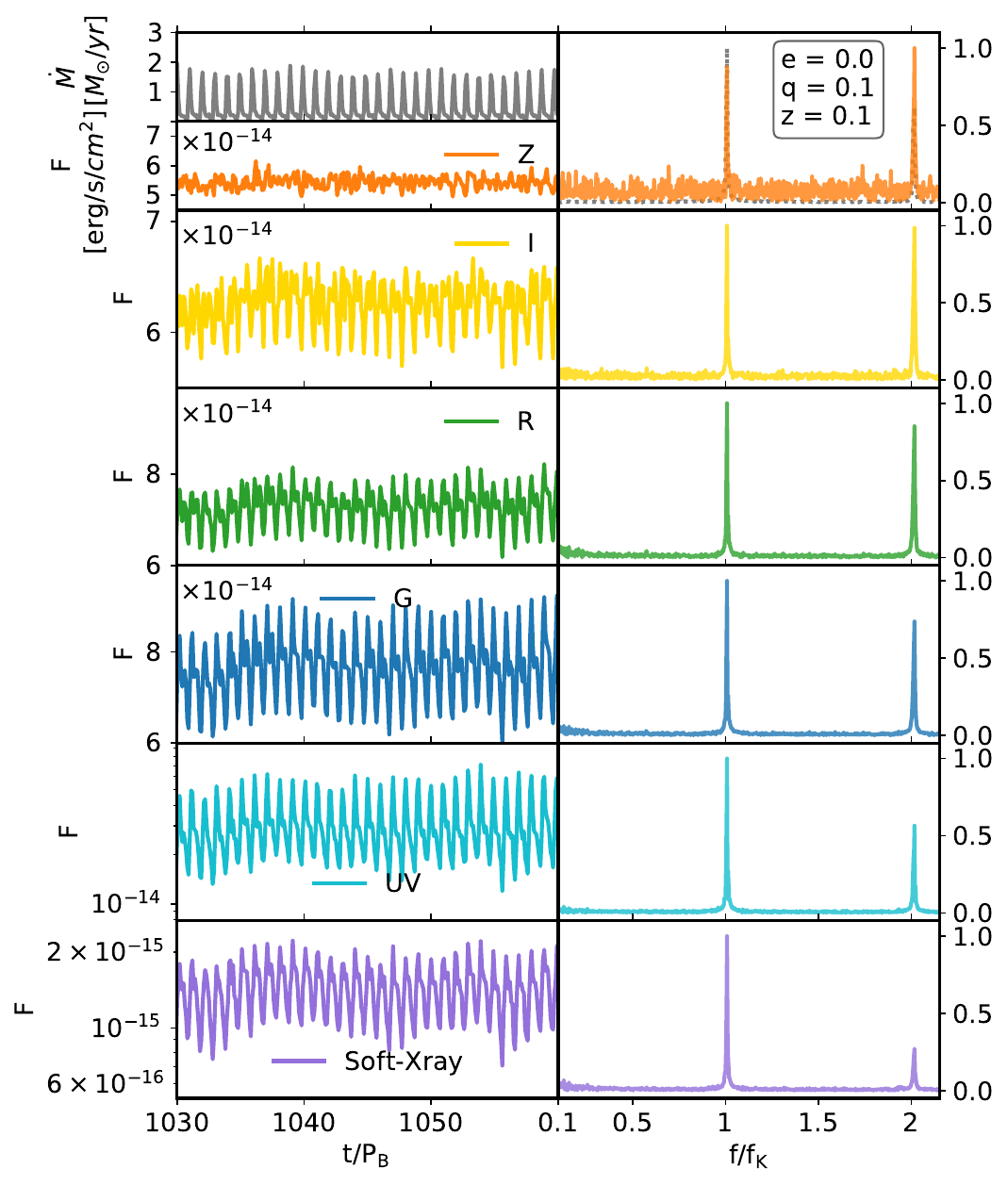}
	\caption{\label{fig:LCse0} Light curves for circular mass binaries at $z = 0.1$.  The left and right panels are for binaries with $q=1$ and $q=0.1$ respectively. In each panel, the first row shows the accretion rate (grey line) and the flux (orange line) integrated over the whole frequency range we consider ($10^8-2.8 \times 10^{19} \, \rm {Hz}$). The left column shows the flux evolution in time while the right column shows the FFT of the accretion rate and flux over 300 orbits in the time window $t=1000-1300 \,P_{\rm B}$ (left panel) or $t=760-1060 \,P_{\rm B}$ (right panel), normalised to unity, in function of $\rm f/f_{K}$ with $\rm f_{K}$ the Keplerian frequency of the binary. The second, third and forth row shows the flux and FFT in the optical Z band, UV band and soft X-ray band respectively. The optical flux is computed taking into account an extra Gaussian noise component as described in Section \ref{Emission model}.
 }
\end{figure*}
\subsection{Spectral energy distributions}
\label{Results_SEDsLCs}

We compute the EM emission from our numerical simulations by using Planck's Law (see Eq. \ref{eqn:planck}), taking into account both the gas and the radiation pressure contribution when computing the disc temperature, as mentioned in Section \ref{Emission model}. We divide our simulated domain into five different regions: the mini-discs region that extends from the sink radius of each binary component out to the Roche Lobe size, the streams region that extends from outside the Roche Lobe out to $r = 3a$, the inner $3a < r < 5a $ and outer $5a < r < 10a $ parts of the disc.
For each region, the total SED is obtained by integrating the flux emitted by each pixel over the whole spatial domain.

Figure \ref{fig:SEDe0q1} shows the surface density maps, the effective temperature maps and the SED obtained for each region at $t = 1298 \, P_{\rm B}$ for the circular equal mass binary. Each panel line shows a different orbital phase of the binary. Periodically, a small fraction of gas enters the cavity and feeds the mini-discs around each binary component through the streams.  
We assume an initial disc temperature profile that decreases with radius as $R^{-0.5}$.
The inflow of gas into the cavity combined with the exchange of material among the mini-discs generates shocks that increase the gas temperature, resulting in, an effective temperature $T_{\rm eff} \sim 10^4 \,$K, warmer than the outer parts of disc.

As shown by the four different orbital phases in Figure \ref{fig:SEDe0q1}, the mini-discs and streams temperature variations occur within an orbital period, producing EM emission variations. The spectra obtained by analysing the emission from the mini-discs and the streams region (blue, orange and green line) exhibit indeed more variability than the inner and outer regions of the disc (red and purple line). 
The emission peak of the mini-discs and the streams region changes frequency between the optical and UV band ($\log\,(\nu/\rm{Hz}) \sim 14.6-15.2 $) spanning 1 order of magnitude in luminosity, while the emission peak of the circumbinary disc remains between the IR and optical band during one orbital period.

In highly eccentric $e=0.9$ case (see Figure \ref{fig:SEDe09q1}), the emission from the cavity shows more variability than in the circular case, due to the geometry of the orbit of the components. 
Indeed, the emission peak of the mini-discs and the stream component changes between the optical and UV band ($\log\,(\nu/\rm{Hz}) \sim 14.8-15.4$) spanning 2 orders of magnitude in luminosity ($\log\,(\nu L_{\rm{\nu}}/{\rm erg} \, {\rm s}^{-1}) \sim 39-41$).

The surface density maps, the effective temperature maps and the SEDs are shown in Figures \ref{fig:SEDe0q01} (for $e=0$) and \ref{fig:SEDe09q01} (for $e=0.9$) for unequal mass binaries ($q=0.1$).
We find the mini-discs to have a lower temperature in the  circular unequal mass case with respect to the circular equal mass case shown in Figure \ref{fig:SEDe0q1}. Indeed, the exchange of material between the mini-discs does not produce significant shocks and the temperature does not increase as much as in the circular case. 
Therefore, the emission peak of the mini-discs is shifted to lower frequencies, $\log\,(\nu/\rm{Hz}) \sim 14.8 - 15.2$, and  spans  $\sim $ 2 orders of magnitude in flux within one orbital period. 

In the eccentric unequal mass case (Figure \ref{fig:SEDe09q01}), the mini-discs are significantly depleted as the high eccentricity of the binary causes them to strongly interact with each other and the material is either promptly accreted or flung back to the cavity wall. 
We note that in this case, the variability of the emission from the mini-discs is not totally hidden by the emission from the streams region but does change the spectrum at high frequencies within one orbital period. 

In all simulations, by construction, the emission of the X-ray corona tracks that of the mini-discs as per Eq.~\eqref{eqn:kx}, therefore, depending on the simulation, it can vary by up to two orders of magnitude in luminosity.

\begin{figure*}
    \centering
    \includegraphics[width=\columnwidth]{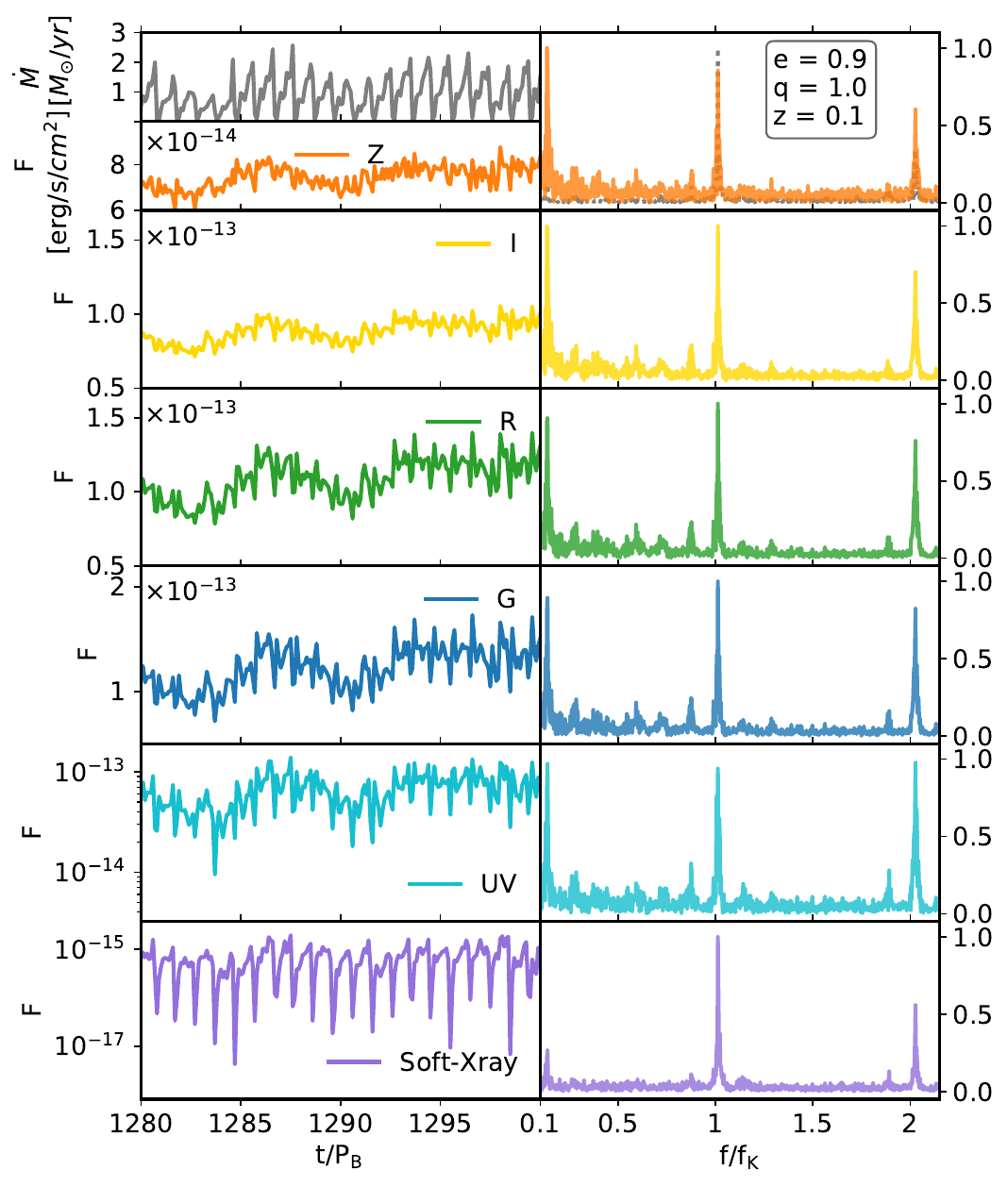}
    \includegraphics[width=\columnwidth]{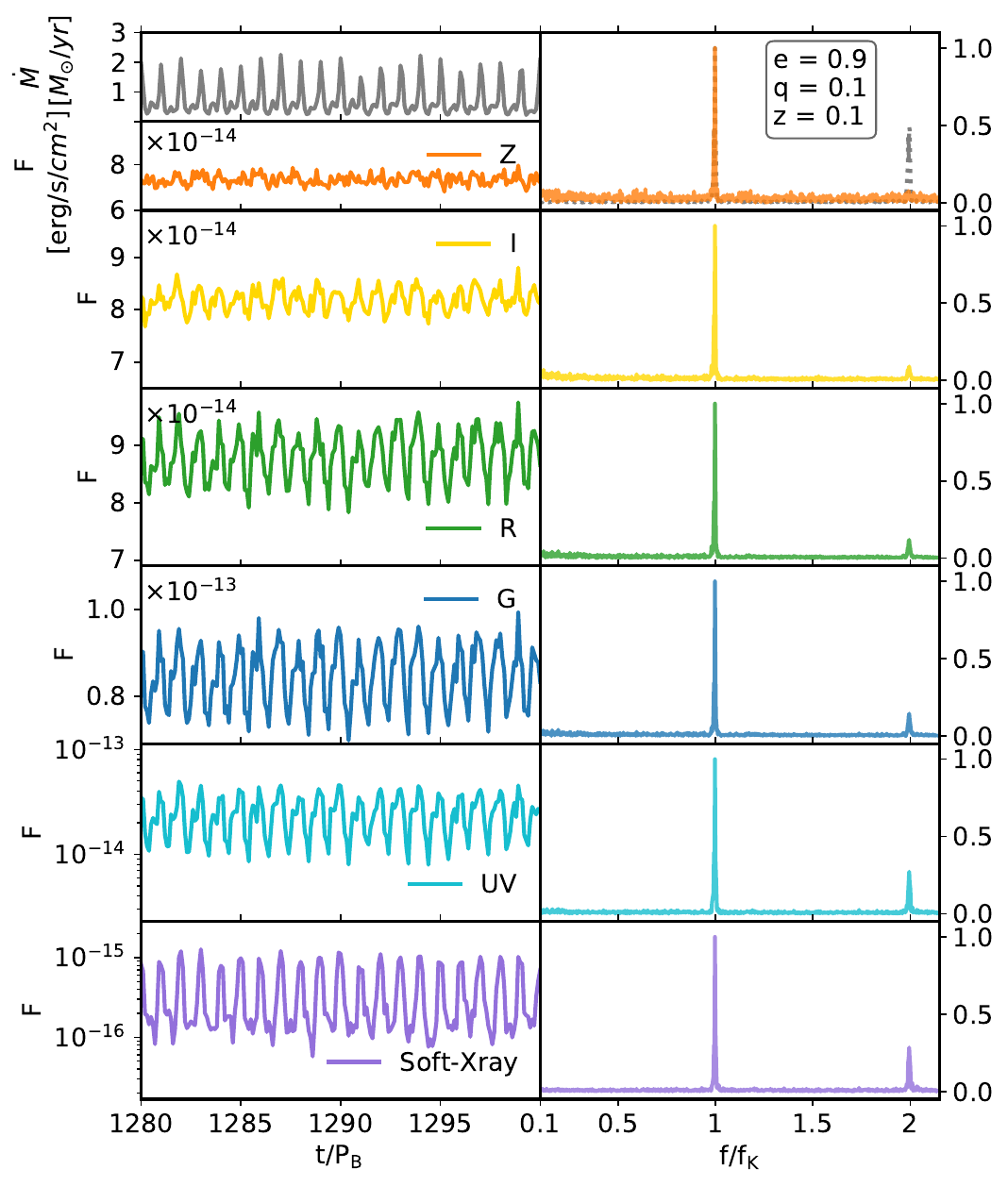}
    \caption{\label{fig:LCse09} 
    Same as Figure \ref{fig:LCse0} but for eccentric $e=0$ binaries. Here, the Fourier Transform of the accretion rate and fluxes is computed over 400 orbits in the time window $t=900-1300 \,P_{\rm B}$. }
\end{figure*}

\subsection{Light curves}
\label{LCs}

We integrate the luminosity emitted in each disc region in order to produce the LCs, showing the flux variation of the system as a function of time. We assume the source to be at redshift $z=0.1$ and investigate the effect of redshift on source observability in the next subsection. 
We have calculated the bolometric flux across the whole frequency range we used to produce the spectra, i.e. $10^8-2.8 \times 10^{19}$ Hz ($4.13 \times 10^{-10} - 41.3$ keV) and in three different region of the EM spectrum: in the optical, using the VRO filters frequency bands (see Table \ref{tab:LSSTbands}), in the near-UV ($1.0-1.5 \times 10^{15}$ Hz or $4.13-6.20$ eV) and in the soft X-ray ($7.25-48.3 \times 10^{16}$ Hz or $0.3-2$ keV).
The upcoming VRO will perform a 10-years survey of the sky in the Southern Hemisphere and it will potentially be able to capture the EM emission from the accretion disc of MBHBs. 
We therefore focus most of our attention to the optical emission of our simulated systems, and we consider the VRO  Z, I, R and G filters (see Table \ref{tab:LSSTbands} for details), and the instrument sensitivity, as described in Section \ref{Emission model}.

Results are shown in Figures \ref{fig:LCse0} ($e=0$) and \ref{fig:LCse09} ($e=0.9$). In each panel of the two figures, the left column shows the LCs computed in all the considered  bands together with the accretion rate, (grey line in the top panel), while the right column shows the Fast Fourier Transform (FFT) of the accretion rate and of the fluxes computed over 300 orbits, $t = 1000-1300 \, P_{\rm B }$ ($t = 760-1060\, P_{\rm B }$ for the case $q=0.1$, $e=0$), normalised to unity. In the VRO filters, the flux is computed including extra Gaussian stochastic fluctuations mimicking the effect of the VRO sensitivity, as described in section \ref{Emission model}. In all cases we see that the emission is brighter in the optical and UV band (except in the eccentric unequal case), while it is dimmer in the soft X-ray, consistent with our SEDs (see Figures \ref{fig:SEDe0q1}--\ref{fig:SEDe09q01}). 
Indeed, in our model the only contribution that produces luminosity at high frequencies is the corona. Another clear trend shown in all simulations is that variability tends to increase with frequency. In fact, while the flux in the VRO bands changes within a factor 2-3, the UV and X-ray fluxes can experience oscillations of more than one order of magnitude, in particular in eccentric cases. This is consistent with the physics of the emission from the disc. The optical mostly comes from the circumbinary disc, which is relatively steady and is only mildly affected by the binary. Conversely the UV emission is dominated by the streams and mini-discs, which are strongly impacted by the dynamics of the binary and thus highly variable. Finally the X-ray corona is directly connected to the UV emission of the mini-discs, which is the component showing the highest variability.

The left panel in Figure \ref{fig:LCse0} displays a number of interesting features. It is  clear that the main modulation pattern is not related to the binary period, but occurs on longer timescales, and this is true bot for the LCs and for the accretion rate. This is confirmed by the FFTs which show two clear peaks at $0.2\,f_{\rm K}$ and $0.4\,f_{\rm K}$ (second harmonic), where $f_{\rm K}$ is the Keplerian frequency of the binary. This periodicity is associated to the "lump", an over-density region that obits at the cavity edge with an orbital period a few times the binary orbital period, as reported in previous works \citep{Macfadyen2008, Cuadra2009, Krolik2010, roedig2011, Noble2012, Shi2012, dorazio2013, farris2014, Bowen2018, Tang2018, Westernacher2022, Westernacher2023}. It is also interesting to notice that, while the accretion rate shows an intricate structure of harmonics \citep{farris2014,Munoz2019,Franchini2023}, this is much less evident in the LCs, where there is significant power only at $2\,f_{\rm K}$, which corresponds to one-half of the orbital period. 
The situation is strikingly different when the mass ratio of the binary is small (right panel in Figure \ref{fig:LCse0}). In this case both the accretion rate and the LCs show a clear periodicity on the binary period, which is confirmed by the FFT, where clear peaks are visible at $1\,f_{\rm K}$ and $2\,f_{\rm K}$ (second harmonic). Also note that there is no clear power at $f<f_{\rm K}$. This is because no significant lump forms in this case, since perturbation induced by the binary are not sufficient to excite an $m=1$ mode at the inner edge of the circumbinary disc cavity. This is in line with results of 2D simulations in the literature \citep[see e.g.][]{farris2014}.

\begin{figure*}
	\begin{center}
    \includegraphics[width=18cm]{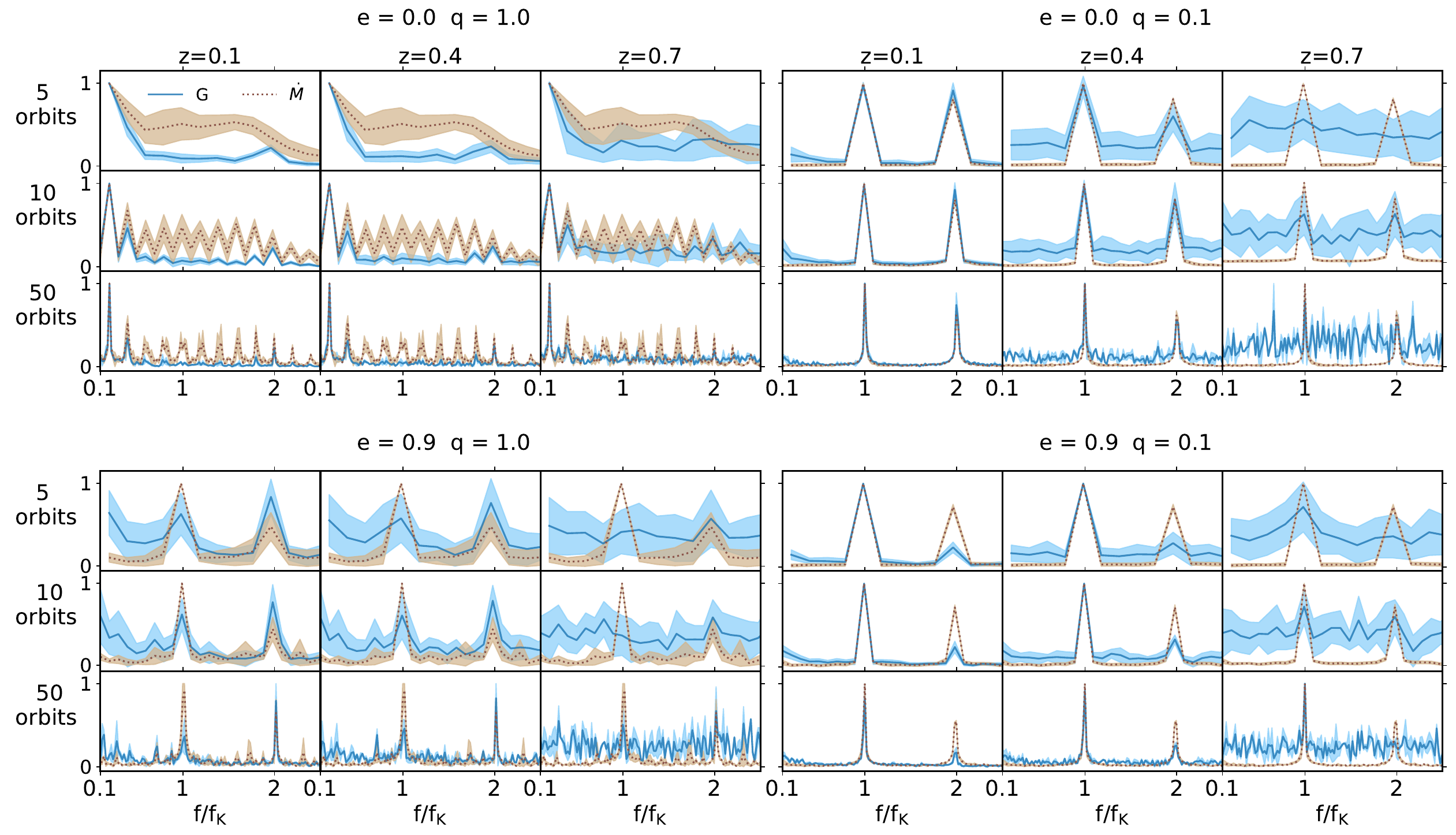}
    \caption{Fast Fourier Transform of Light Curves from the simulation of circular (top) and eccentric $e=0.9$ binaries (bottom) with mass ratio $q=1$ (left column) and $q=0.1$ (right column). The first row of each case shows the FFT of the Optical G band flux (blue line) and the FFT of the accretion rate (brown line) computed over 5-orbits windows within a total of 300/400 orbital periods at redshift $\rm{z=0.1,\,0.4,\,0.7}$. The second and the third rows show the FFT computed over 10- and 50-orbits windows, respectively. 
}
    \label{fig:FFTs_comparison_Ltrue}
	\end{center}
\end{figure*}

Results for the eccentric binary simulations are shown in Figure \ref{fig:LCse09}. The equal mass case (left panel) shows an interesting periodicity structure. The flux emitted in the optical bands all exhibit a clear modulation of a factor $\approx2$, combining periodicities related to both the binary and the lump dynamics. In the FFT, we can clearly see the lump periodicity at $f\approx 0.15f_{\rm K}$: this is lower than the circular binary case, as the cavity is larger and the period associated with its inner edge is $\propto R^{3/2}$. Contrary to the circular equal mass case, clear peaks are visible also at $1\,f_{\rm K}$ and $2\,f_{\rm K}$ (second harmonic), driven by the binary orbital period. 
In the eccentric unequal mass case instead (right panel), the lump periodicity is absent and the peak at $f\sim 2 f_{\rm K}$ is less prominent than what found in the circular unequal binary case.
We have also explored the effect of placing the binary at different redshifts. For simplicity, we discuss the main results without including plots. Besides the obvious difference in flux, making closer binaries easier to detect, there is also some minor change difference in the displayed periodicities due to the redshifting of the spectra at $z<\,0.6$. At redshifts $z>\,0.6$, fluxes are very noisy and periodicities peaks are not always distinguishable in all the optical bands. However, all the main features described here for binaries at $z=0.1$ remain valid.

\begin{figure*}
	\begin{center}
    \includegraphics[width=18cm]{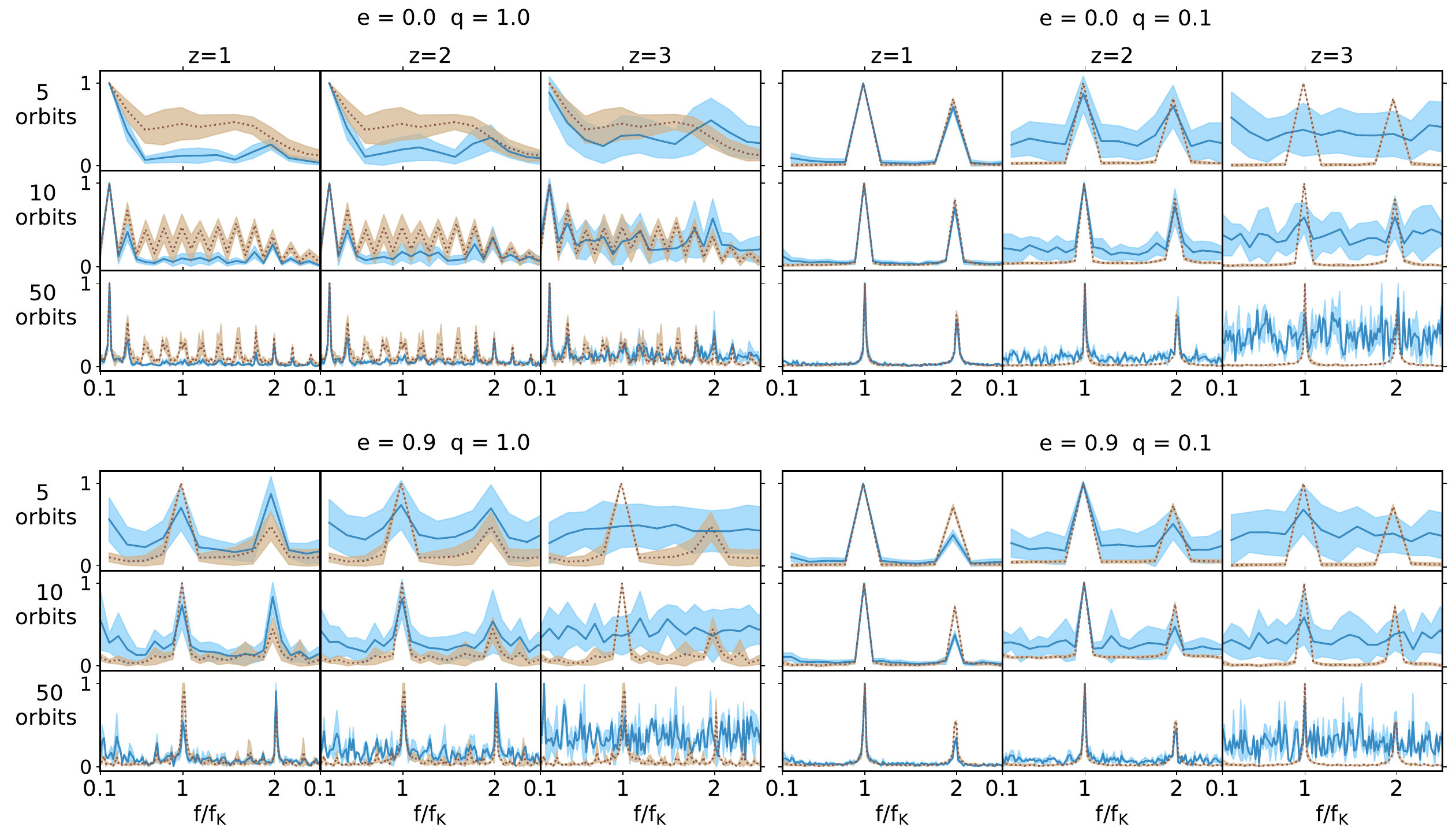}
    \caption{Same as Figure \ref{fig:FFTs_comparison_Ltrue} but assuming that the emitted luminosity is 100 times higher than the luminosity we obtained from simulations. 
    }
    \label{fig:FFTs_comparison_L100}
	\end{center}
\end{figure*}


\subsection{Periodicity identification in the VRO survey}

We have so far computed the FFT over a large number of orbits in order to distinguish the different periodicities associated with binaries with different mass ratio and eccentricity. However, VRO might only have a handful of binary orbits at its disposal in its 10 years observational campaign. This is because binaries with periods shorter than a few years are expected to be primarily driven by GW emission, meaning that wide binaries (with long periods) live longer and are more likely to be present in the data \citep[see][]{Kelly2019}. In fact, the binary period of 1 year considered here was chosen to be representative of the typical system that might be detected with VRO.  
To get a qualitative idea of whether the periodicities found in the previous section can be identified in VRO data, we have computed the FFT over 5, 10 and 50 binary orbits along a total of 300 orbits for the circular binary cases and 400 orbits for the eccentric binaries, essentially shifting the FFT window within these few hundred orbits. We have then computed the average FFT and its standard deviation (STD) at different redshifts, $z=0.1,\,0.4,\,0.7$.

In Figure \ref{fig:FFTs_comparison_Ltrue}, 
we show the results for each of the four binaries we considered. In each plot, the first row shows the mean FFT both of the flux in the optical G band (blue line) and of the accretion rate (brown line) using a 5-orbits window at redshift $z=0.1,\,0.4,\,0.7$ from the left to the right. The second and the third rows show the results obtained with 10- and 50-orbits windows, respectively. Note that, given a VRO survey of 10 years, 5, 10 and 50 orbits correspond to binaries with period of two years, one year and 2.4 months respectively. 

The main result of this exercise is that periodicities are more easily observed in unequal mass binaries than in equal mass ones. In fact, starting from the top left panel of Figure \ref{fig:FFTs_comparison_Ltrue}, we see that periodic features in circular equal mass binaries will be extremely hard to pick. With only 5 orbits, the lump periodicity falls in the lowest frequency bin of the Fourier decomposition and cannot be securely identified. Other periodicities are much weaker and do not show significant power in the FFT over 5 orbits. Things improve with increasing numbers of orbits and the lump periodicity clearly emerges when considering 50 orbits.
Conversely, the circular unequal mass case (top right panel) shows that the peaks at $f=f_{\rm K}$ and $f=2f_{\rm K}$ are already prominent after only 5 orbits, which is very promising, and the situation naturally improves if more orbits are sampled.

A similar situation is observed for eccentric binaries, with periodicities that are more prominent in the unequal mass case. There are noticeable differences though. In the equal mass case (lower left panel), periodicities at $f=f_{\rm K}$ and $f=2f_{\rm K}$ start to emerge already after 5 orbits. However, the variance is large, meaning that these features are not prominent and might be hard to detect. The situation naturally improves with the number of orbits, but a large variance remains, even after 50 orbits. It is also interesting to note that the lump frequency, clearly present in the circular case, does not seem to emerge here. In the eccentric unequal mass case (lower right panel), the orbital periodicity is clearly identified already after 5 orbits, although the second harmonic is much less prominent than in the circular case.

In all cases, the variance tends to increase with redshift. This is a natural effect due to the inclusion of the VRO sensitivity limit in the computation. In particular, the VRO telescope could encounter substantial difficulties in detecting flux periodicities emanating from binaries with mass $M_{\rm B} \sim 10^6 M_{\odot}$ at redshift $z > \, 0.5$. This is clearly illustrated by the $z=0.7$ panels of Figure \ref{fig:FFTs_comparison_Ltrue}, where the peaks generally observable at $z=0.1, 0.4$ tend to be swamped in the variance.

This does not necessarily mean that VRO cannot identify LISA MBHB precursors beyond $z\approx 0.5$. In fact, the bolometric luminosity of the systems simulated here is $L_{\rm bol}\approx 10^{42}$ erg s$^{-1}\approx 0.01 L_{\rm Edd}$. The largely sub-Eddington luminosity is mainly due to the relatively low temperature of the gas, reaching only $T\approx 3\times 10^4$K, which is a factor of a few cooler than the temperature of a standard thin disk around a $10^6$M$_\odot$ MBH \citep{2002apa..book.....F}. This is also expected, due to the fact that $R_{\rm sink}=0.05a \sim 100 R_{\rm ISCO}$, as mentioned in Section \ref{discmodel}, and the overall luminosity of the system might indeed be higher.
We therefore explore here also the detectability of a putative brighter binary by simply multiplying the emission by a factor of 100, i.e. $L=100 \times L_{\rm true}$, thus preserving all the variability properties found in our simulations. Although this is not self-consistent with our simulation, we consider it a useful exercise to assess VRO performances against lightcurves that could be representative of an $M=10^6$M$_\odot$ MBHBs emitting at the Eddington luminosity, or potentially of a more massive system of $M=10^8$M$_\odot$ but emitting at $\approx 0.01 L_{\rm Edd}$.
We repeated the process described above: we analyse the FFT of the flux over 5, 10 and 50 binary orbits along a total of 300 orbits for the circular binary cases and 400 orbits for the eccentric binaries and we compute the average FFT and its STD for different redshifts $z =\, 1, \,2, \,3$. Results are shown in Figure \ref{fig:FFTs_comparison_L100}. Most of the features discussed for the case $L=L_{\rm true}$ are still observed, but now peaks in the FFT can be easily identified up to $z=2$, around cosmic noon. A full statistical assessment of VRO capabilities of correctly identifying these peaks will be the subject of future work. 



\section{Conclusions}
\label{sec:conclusions}

In this work, we computed SEDs and multi-wavelength LCs from 3D numerical simulations with hyper-Lagrangian refinement of milliparsec scale MBHBs embedded in thin circumbinary gaseous discs. The discs in our simulations are described by an adiabatic equation of state. We therefore allowed the gas to heat due to shocks, viscosity, and PdV work, and to cool via black body radiation.
We explored binary eccentricities $e=0, \, 0.9$ and mass ratio $q=1,\, 0.1$, and computed LCs in different bands placing the sources at different redshifts, $z=0.1, \,0.4, \, 0.7$. 

We investigated the evolution of the disc aspect ratio $H/R$, the binary semi-major axis $a$ and the binary eccentricity $e$. We found that, after an initial phase where the black body cooling dominates the gas thermodynamic evolution, the disc experiences thickens again, reaching $H/R \sim 0.08$ in the inner parts and maintaining a lower $ H/R \sim 0.04$ (corresponding to a lower temperature) in the outer part of the disc, regardless of the initial choice of binary mass ratio and eccentricity. The final equilibrium state is, in fact, mostly driven by initial disc mass and radial extension, which are the same in all simulations. 
Therefore self-regulation \citep{lodato2007sg} drives all discs to reach a similar aspect ratio at the end of the transient phase.
We find the interaction between the binary and the circumbinary disc to cause the binary to shrink, regardless of the initial conditions. Since in our model the temperature changes with time, this result further supports previous findings in the literature \citep{roedig2014,Franchini2021}. 
We find that circular orbits tend towards higher eccentricity values, in agreement with previous works \citep[e.g.][]{DOrazio2021, Siwek2023}, whereas very eccentric binaries experience a negligible eccentricity evolution within the timeframe of our simulations. We notice, however, that we followed the evolution of the relaxed disc only for about 400 orbits, corresponding to 400 years. It might be that the eccentricity evolution for very eccentric binaries occurs on longer timescale than what simulated here. Indeed, here we simulated the binary evolution over $\sim 1300 \, yr$ while in both the cited works the timescale is longer.

We computed the SEDs from the circumbinary discs in our simulations assuming black body emission.
We find that the luminosity emitted by the innermost region of the disc, i.e. the mini-discs and the streams, exhibits more variability than the outer parts of the disc.
In the circular equal mass case, the emission peak of the mini-discs and the streams region changes frequency between the optical and UV band ($\log\,(\nu/\rm{Hz}) \sim 14.8 - 15.4$) spanning 1 order of magnitude in luminosity. In the unequal mass case the mini-discs have a lower temperature due to the absence of shocks produced by the exchange of material among the binary components. Thus, the emission peak of the mini-discs is shifted to slightly lower frequencies i.e. $\log\,(\nu/\rm{Hz}) \sim 14.8-15.2$, and spans up to $\sim 2 $ orders of magnitude in luminosity. 
In both the eccentric binary cases, we find that the emission from the mini-discs is completely (in the equal mass case) or partially (in the unequal mass case) covered by the streams emission. 
The X-ray photons are provided by the corona, that we assume to have an emission proportional to that of the mini-discs. Therefore, the luminosity in the X-ray band displays the highest variability, which can reach up to two orders of magnitude for unequal mass binaries.

We computed the LCs in different frequency bands, mainly focusing on the optical window that will be probed by VRO. We calculated the (thermal) flux emitted over the whole EM spectrum that we used to produce the SEDs, i.e. within the frequency band $10^8-2.8 \times 10^{19}$ Hz  ($4.13 \times 10^{-10}-41.3$ keV), in the optical frequency band using the VRO filters, in the near-UV band within $1.0 - 1.5 \times 10^{15}$ Hz ($4.13-6.20$ eV) and in the soft X-ray band, i.e. in the range $7.25-48.3 \times 10^{16}$ Hz ($0.3- 2$ keV).

In almost all the cases, the flux is notably higher in the UV band, while in the soft X-ray is dimmer, consistent with the shape of the SEDs. As the frequency increases the flux variability grows, in particular in the UV and Soft X-ray bands. Here the flux oscillates by more than  an order of magnitude while in the VRO optical bands fluxes vary within a factor of two, in line with the physics of emission from the disc.

In the circular equal mass case, both the flux and the accretion rate FFTs reveal clear peaks at at $0.2 \, f_{\rm K}$ and  $0.4 \, f_{\rm K}$ associated to the lump periodicity. Moreover, in the LCs, a peak at $2 \, f_{\rm K}$ is also present, which corresponds to a periodicity of one-half the binary orbital period. 
In the eccentric equal mass case, the lump periodicity is significant only in the optical and UV fluxes, while its amplitude is negligible in the soft X-ray band. This is probably due to the larger cavity carved by the binary that causes the lump region to emit in the optical/UV band rather than in the soft X-ray band.  
We indeed found the lump modulation peak to be shifted to $f_{\rm K} \sim 0.15$. This is consistent with the fact that the cavity is larger than in the circular equal mass case since the period associated with its inner edge is $\propto R^{3/2}$.
We note that we found evidence of lump periodicity only in equal mass binaries. This is consistent with previous works that show the lump modulation amplitude to decrease with binary mass ratio \citep{dorazio2013}.
Therefore a lack of sub-orbital modulation in the presence of a clear orbital modulation might indicate a small binary mass ratio.
We also found a prominent flux and accretion rate modulation over the orbital period of the binary and half of it in all the simulations, with the exception of the circular equal mass case, which shows a a weak periodicity only at $2\, f_{\rm K}$.
By exploring different redshifts, there are some minor change in periodicities at redshift $z<\, 0.6$, while at higher redshifts fluxes are very noisy and the periodicities are not always well distinguished.

All the aforementioned considerations are valid when the FFT is computed over a large number of binary orbits (i.e. $300-400 \, P_{\rm B}$). However, the VRO survey is planned for 10 years and most compact MBHBs are expected to have periods of $\approx$years. 
Therefore, we have computed the FFT of the flux and of the accretion rate over 5, 10 and 50 binary obits at different redshifts to make more realistic assessment of the prominence of these periodicities in VRO data. 
In the circular equal mass case, detecting periodicities is challenging, in particular with only 5 orbits. We found that, as the number of orbits used to compute the FFT increases, the periodicities become more distinct and the associated variance decreases, as expected. Still, identification of equal mass, circular binaries appears to be the most challenging.
Conversely, binaries with $q=0.1$ show promising results, with distinguishable peaks at $1\, f_{\rm K}$ and $2 \, f_{\rm K}$ even after 5 orbits. 
Similar trends are observed in eccentric cases. 
The lump periodicity is totally absent in all the cases but for the circular equal mass case, that shows a hint of lump periodicity when computing the FFT over 50 orbits. Thus, the chances of detecting it during a 10-year survey by assuming unequal MBHB with orbital period of 1 year are negligible.

Due to the intrinsic faintness of our system, $L_{\rm bol}\approx 10^{42}$ erg s$^{-1}$, the detectability of periodicities with VRO are limited to systems at $z<0.5$.
As an exercise, we increased the luminosity of all our systems by a factor of 100, mimicking a MBHB of $M=10^6$M$_\odot$ emitting at the Eddington limit. We found that in this case, periodicities can be identified by VRO up to $z\approx 2$, opening the possibility to find these systems in a large fraction of the co-moving volume of the Universe.

In general, our results indicate that periodicities related to unequal mass binaries will be easier to identify in VRO data, compared to equal mass ones. Moreover, LCs from binaries with different properties (equal vs unequal mass, circular vs eccentric) are characterised by different distinctive modulations, hinting to the possibility (to be further investigated) of constraining the binary properties from their time-domain EM data. 

In this work, we present advanced 3D simulations that improve the description of accreting MBHBs embedded in gaseous discs, which is fundamental in order to make more realistic predictions of the emission signatures of these sources.
We note here that we included the effect that radiation pressure has in determining the gas temperature only a posteriori. The inclusion of radiation pressure on the fly is the subject of a future work (Cocchiararo et al. in prep). The inclusion of radiation pressure will allow us to obtain a more comprehensive model of these systems and possibly a better characterisation of their EM signatures. Finally, a wider exploration of the binary-disc parameter space using our numerical simulations is needed to make more robust observational predictions and will be the subject of future work.

\section*{Acknowledgements}

We thank Daniel Price for providing the {\sc phantom} code for numerical simulations and acknowledge the use of {\sc splash} \citep{Price2007} for the rendering of the figures.
We thank Phil Hopkins for providing the {\sc gizmo} code for numerical simulations. 
AF and AS acknowledge financial support provided under the European Union’s H2020 ERC Consolidator Grant ``Binary Massive Black Hole Astrophysics" (B Massive, Grant Agreement: 818691). AL acknowledges support by the PRIN MUR "2022935STW".
FC thanks Nataliya Porayko, Golam Shaifullah and Enrico Panontin for helpful discussions and suggestions.


%
%
\bibliographystyle{aa} 
\bibliography{bibliography}

\begin{thebibliography}{99}
\expandafter\ifx\csname natexlab\endcsname\relax\def\natexlab#1{#1}\fi

\bibitem[{{Afzal} {et~al.}(2023){Afzal}, {Agazie}, {Anumarlapudi}, {Archibald},
  {Arzoumanian}, {Baker}, {B{\'e}csy}, {Blanco-Pillado}, {Blecha}, {Boddy},
  {Brazier}, {Brook}, {Burke-Spolaor}, {Burnette}, {Case}, {Charisi},
  {Chatterjee}, {Chatziioannou}, {Cheeseboro}, {Chen}, {Cohen}, {Cordes},
  {Cornish}, {Crawford}, {Cromartie}, {Crowter}, {Cutler}, {Decesar}, {Degan},
  {Demorest}, {Deng}, {Dolch}, {Drachler}, {von Eckardstein}, {Ferrara},
  {Fiore}, {Fonseca}, {Freedman}, {Garver-Daniels}, {Gentile}, {Gersbach},
  {Glaser}, {Good}, {Guertin}, {G{\"u}ltekin}, {Hazboun}, {Hourihane}, {Islo},
  {Jennings}, {Johnson}, {Jones}, {Kaiser}, {Kaplan}, {Kelley}, {Kerr}, {Key},
  {Laal}, {Lam}, {Lamb}, {Lazio}, {Lee}, {Lewandowska}, {Lino Dos Santos},
  {Littenberg}, {Liu}, {Lorimer}, {Luo}, {Lynch}, {Ma}, {Madison}, {McEwen},
  {McKee}, {McLaughlin}, {McMann}, {Meyers}, {Meyers}, {Mingarelli},
  {Mitridate}, {Nay}, {Natarajan}, {Ng}, {Nice}, {Ocker}, {Olum}, {Pennucci},
  {Perera}, {Petrov}, {Pol}, {Radovan}, {Ransom}, {Ray}, {Romano}, {Sardesai},
  {Schmiedekamp}, {Schmiedekamp}, {Schmitz}, {Schr{\"o}der}, {Schult},
  {Shapiro-Albert}, {Siemens}, {Simon}, {Siwek}, {Stairs}, {Stinebring},
  {Stovall}, {Stratmann}, {Sun}, {Susobhanan}, {Swiggum}, {Taylor}, {Taylor},
  {Trickle}, {Turner}, {Unal}, {Vallisneri}, {Verma}, {Vigeland}, {Wahl},
  {Wang}, {Witt}, {Wright}, {Young}, {Zurek}, \& {Nanograv
  Collaboration}}]{2023ApJ...951L..11A}
{Afzal}, A., {Agazie}, G., {Anumarlapudi}, A., {et~al.} 2023, \apjl, 951, L11

\bibitem[{{Agazie} {et~al.}(2023{\natexlab{a}}){Agazie}, {Alam},
  {Anumarlapudi}, {Archibald}, {Arzoumanian}, {Baker}, {Blecha}, {Bonidie},
  {Brazier}, {Brook}, {Burke-Spolaor}, {B{\'e}csy}, {Chapman}, {Charisi},
  {Chatterjee}, {Cohen}, {Cordes}, {Cornish}, {Crawford}, {Cromartie},
  {Crowter}, {Decesar}, {Demorest}, {Dolch}, {Drachler}, {Ferrara}, {Fiore},
  {Fonseca}, {Freedman}, {Garver-Daniels}, {Gentile}, {Glaser}, {Good},
  {G{\"u}ltekin}, {Hazboun}, {Jennings}, {Jessup}, {Johnson}, {Jones},
  {Kaiser}, {Kaplan}, {Kelley}, {Kerr}, {Key}, {Kuske}, {Laal}, {Lam}, {Lamb},
  {Lazio}, {Lewandowska}, {Lin}, {Liu}, {Lorimer}, {Luo}, {Lynch}, {Ma},
  {Madison}, {Maraccini}, {McEwen}, {McKee}, {McLaughlin}, {McMann}, {Meyers},
  {Mingarelli}, {Mitridate}, {Ng}, {Nice}, {Ocker}, {Olum}, {Panciu},
  {Pennucci}, {Perera}, {Pol}, {Radovan}, {Ransom}, {Ray}, {Romano}, {Salo},
  {Sardesai}, {Schmiedekamp}, {Schmiedekamp}, {Schmitz}, {Shapiro-Albert},
  {Siemens}, {Simon}, {Siwek}, {Stairs}, {Stinebring}, {Stovall}, {Susobhanan},
  {Swiggum}, {Taylor}, {Turner}, {Unal}, {Vallisneri}, {Vigeland}, {Wahl},
  {Wang}, {Witt}, {Young}, \& {Nanograv Collaboration}}]{2023ApJ...951L...9A}
{Agazie}, G., {Alam}, M.~F., {Anumarlapudi}, A., {et~al.} 2023{\natexlab{a}},
  \apjl, 951, L9

\bibitem[{{Agazie} {et~al.}(2023{\natexlab{b}}){Agazie}, {Anumarlapudi},
  {Archibald}, {Arzoumanian}, {Baker}, {B{\'e}csy}, {Blecha}, {Brazier},
  {Brook}, {Burke-Spolaor}, {Burnette}, {Case}, {Charisi}, {Chatterjee},
  {Chatziioannou}, {Cheeseboro}, {Chen}, {Cohen}, {Cordes}, {Cornish},
  {Crawford}, {Cromartie}, {Crowter}, {Cutler}, {Decesar}, {Degan}, {Demorest},
  {Deng}, {Dolch}, {Drachler}, {Ellis}, {Ferrara}, {Fiore}, {Fonseca},
  {Freedman}, {Garver-Daniels}, {Gentile}, {Gersbach}, {Glaser}, {Good},
  {G{\"u}ltekin}, {Hazboun}, {Hourihane}, {Islo}, {Jennings}, {Johnson},
  {Jones}, {Kaiser}, {Kaplan}, {Kelley}, {Kerr}, {Key}, {Klein}, {Laal}, {Lam},
  {Lamb}, {Lazio}, {Lewandowska}, {Littenberg}, {Liu}, {Lommen}, {Lorimer},
  {Luo}, {Lynch}, {Ma}, {Madison}, {Mattson}, {McEwen}, {McKee}, {McLaughlin},
  {McMann}, {Meyers}, {Meyers}, {Mingarelli}, {Mitridate}, {Natarajan}, {Ng},
  {Nice}, {Ocker}, {Olum}, {Pennucci}, {Perera}, {Petrov}, {Pol}, {Radovan},
  {Ransom}, {Ray}, {Romano}, {Sardesai}, {Schmiedekamp}, {Schmiedekamp},
  {Schmitz}, {Schult}, {Shapiro-Albert}, {Siemens}, {Simon}, {Siwek}, {Stairs},
  {Stinebring}, {Stovall}, {Sun}, {Susobhanan}, {Swiggum}, {Taylor}, {Taylor},
  {Turner}, {Unal}, {Vallisneri}, {van Haasteren}, {Vigeland}, {Wahl}, {Wang},
  {Witt}, {Young}, \& {Nanograv Collaboration}}]{nanograv2023}
{Agazie}, G., {Anumarlapudi}, A., {Archibald}, A.~M., {et~al.}
  2023{\natexlab{b}}, \apjl, 951, L8

\bibitem[{{Agazie} {et~al.}(2023{\natexlab{c}}){Agazie}, {Anumarlapudi},
  {Archibald}, {Arzoumanian}, {Baker}, {B{\'e}csy}, {Blecha}, {Brazier},
  {Brook}, {Burke-Spolaor}, {Burnette}, {Case}, {Charisi}, {Chatterjee},
  {Chatziioannou}, {Cheeseboro}, {Chen}, {Cohen}, {Cordes}, {Cornish},
  {Crawford}, {Cromartie}, {Crowter}, {Cutler}, {Decesar}, {Degan}, {Demorest},
  {Deng}, {Dolch}, {Drachler}, {Ellis}, {Ferrara}, {Fiore}, {Fonseca},
  {Freedman}, {Garver-Daniels}, {Gentile}, {Gersbach}, {Glaser}, {Good},
  {G{\"u}ltekin}, {Hazboun}, {Hourihane}, {Islo}, {Jennings}, {Johnson},
  {Jones}, {Kaiser}, {Kaplan}, {Kelley}, {Kerr}, {Key}, {Klein}, {Laal}, {Lam},
  {Lamb}, {Lazio}, {Lewandowska}, {Littenberg}, {Liu}, {Lommen}, {Lorimer},
  {Luo}, {Lynch}, {Ma}, {Madison}, {Mattson}, {McEwen}, {McKee}, {McLaughlin},
  {McMann}, {Meyers}, {Meyers}, {Mingarelli}, {Mitridate}, {Natarajan}, {Ng},
  {Nice}, {Ocker}, {Olum}, {Pennucci}, {Perera}, {Petrov}, {Pol}, {Radovan},
  {Ransom}, {Ray}, {Romano}, {Sardesai}, {Schmiedekamp}, {Schmiedekamp},
  {Schmitz}, {Schult}, {Shapiro-Albert}, {Siemens}, {Simon}, {Siwek}, {Stairs},
  {Stinebring}, {Stovall}, {Sun}, {Susobhanan}, {Swiggum}, {Taylor}, {Taylor},
  {Turner}, {Unal}, {Vallisneri}, {van Haasteren}, {Vigeland}, {Wahl}, {Wang},
  {Witt}, {Young}, \& {Nanograv Collaboration}}]{2023ApJ...951L...8A}
{Agazie}, G., {Anumarlapudi}, A., {Archibald}, A.~M., {et~al.}
  2023{\natexlab{c}}, \apjl, 951, L8

\bibitem[{{Agazie} {et~al.}(2023{\natexlab{d}}){Agazie}, {Anumarlapudi},
  {Archibald}, {Arzoumanian}, {Baker}, {B{\'e}csy}, {Blecha}, {Brazier},
  {Brook}, {Burke-Spolaor}, {Charisi}, {Chatterjee}, {Cohen}, {Cordes},
  {Cornish}, {Crawford}, {Cromartie}, {Crowter}, {Decesar}, {Demorest},
  {Dolch}, {Drachler}, {Ferrara}, {Fiore}, {Fonseca}, {Freedman},
  {Garver-Daniels}, {Gentile}, {Glaser}, {Good}, {Guertin}, {G{\"u}ltekin},
  {Hazboun}, {Jennings}, {Johnson}, {Jones}, {Kaiser}, {Kaplan}, {Kelley},
  {Kerr}, {Key}, {Laal}, {Lam}, {Lamb}, {Lazio}, {Lewandowska}, {Liu},
  {Lorimer}, {Luo}, {Lynch}, {Ma}, {Madison}, {McEwen}, {McKee}, {McLaughlin},
  {McMann}, {Meyers}, {Mingarelli}, {Mitridate}, {Ng}, {Nice}, {Ocker}, {Olum},
  {Pennucci}, {Perera}, {Pol}, {Radovan}, {Ransom}, {Ray}, {Romano},
  {Sardesai}, {Schmiedekamp}, {Schmiedekamp}, {Schmitz}, {Shapiro-Albert},
  {Siemens}, {Simon}, {Siwek}, {Stairs}, {Stinebring}, {Stovall}, {Susobhanan},
  {Swiggum}, {Taylor}, {Turner}, {Unal}, {Vallisneri}, {Vigeland}, {Wahl},
  {Witt}, {Young}, \& {Nanograv Collaboration}}]{2023ApJ...951L..10A}
{Agazie}, G., {Anumarlapudi}, A., {Archibald}, A.~M., {et~al.}
  2023{\natexlab{d}}, \apjl, 951, L10

\bibitem[{{Amaro-Seoane} {et~al.}(2023){Amaro-Seoane}, {Andrews}, {Arca Sedda},
  {Askar}, {Baghi}, {Balasov}, {Bartos}, {Bavera}, {Bellovary}, {Berry},
  {Berti}, {Bianchi}, {Blecha}, {Blondin}, {Bogdanovi{\'c}}, {Boissier},
  {Bonetti}, {Bonoli}, {Bortolas}, {Breivik}, {Capelo}, {Caramete},
  {Cattorini}, {Charisi}, {Chaty}, {Chen}, {Chru{\'s}li{\'n}ska}, {Chua},
  {Church}, {Colpi}, {D'Orazio}, {Danielski}, {Davies}, {Dayal}, {De Rosa},
  {Derdzinski}, {Destounis}, {Dotti}, {Dutan}, {Dvorkin}, {Fabj}, {Foglizzo},
  {Ford}, {Fouvry}, {Franchini}, {Fragos}, {Fryer}, {Gaspari}, {Gerosa},
  {Graziani}, {Groot}, {Habouzit}, {Haggard}, {Haiman}, {Han}, {Istrate},
  {Johansson}, {Khan}, {Kimpson}, {Kokkotas}, {Kong}, {Korol}, {Kremer},
  {Kupfer}, {Lamberts}, {Larson}, {Lau}, {Liu}, {Lloyd-Ronning}, {Lodato},
  {Lupi}, {Ma}, {Maccarone}, {Mandel}, {Mangiagli}, {Mapelli}, {Mathis},
  {Mayer}, {McGee}, {McKernan}, {Miller}, {Mota}, {Mumpower}, {Nasim},
  {Nelemans}, {Noble}, {Pacucci}, {Panessa}, {Paschalidis}, {Pfister},
  {Porquet}, {Quenby}, {Ricarte}, {R{\"o}pke}, {Regan}, {Rosswog}, {Ruiter},
  {Ruiz}, {Runnoe}, {Schneider}, {Schnittman}, {Secunda}, {Sesana}, {Seto},
  {Shao}, {Shapiro}, {Sopuerta}, {Stone}, {Suvorov}, {Tamanini}, {Tamfal},
  {Tauris}, {Temmink}, {Tomsick}, {Toonen}, {Torres-Orjuela}, {Toscani},
  {Tsokaros}, {Unal}, {V{\'a}zquez-Aceves}, {Valiante}, {van Putten}, {van
  Roestel}, {Vignali}, {Volonteri}, {Wu}, {Younsi}, {Yu}, {Zane}, {Zwick},
  {Antonini}, {Baibhav}, {Barausse}, {Bonilla Rivera}, {Branchesi},
  {Branduardi-Raymont}, {Burdge}, {Chakraborty}, {Cuadra}, {Dage}, {Davis}, {de
  Mink}, {Decarli}, {Doneva}, {Escoffier}, {Gandhi}, {Haardt}, {Lousto},
  {Nissanke}, {Nordhaus}, {O'Shaughnessy}, {Portegies Zwart}, {Pound},
  {Schussler}, {Sergijenko}, {Spallicci}, {Vernieri}, \&
  {Vigna-G{\'o}mez}}]{LISA2023}
{Amaro-Seoane}, P., {Andrews}, J., {Arca Sedda}, M., {et~al.} 2023, Living
  Reviews in Relativity, 26, 2

\bibitem[{{Antoniadis} {et~al.}(2023{\natexlab{a}}){Antoniadis}, {Arumugam},
  {Arumugam}, {Auclair}, {Babak}, {Bagchi}, {Bak Nielsen}, {Barausse}, {Bassa},
  {Bathula}, {Berthereau}, {Bonetti}, {Bortolas}, {Brook}, {Burgay},
  {Caballero}, {Caprini}, {Chalumeau}, {Champion}, {Chanlaridis}, {Chen},
  {Cognard}, {Crisostomi}, {Dandapat}, {Deb}, {Desai}, {Desvignes},
  {Dhanda-Batra}, {Dwivedi}, {Falxa}, {Fastidio}, {Ferdman}, {Franchini},
  {Gair}, {Goncharov}, {Gopakumar}, {Graikou}, {Grie{\ss}meier}, {Gualandris},
  {Guillemot}, {Guo}, {Gupta}, {Hisano}, {Hu}, {Iraci}, {Izquierdo-Villalba},
  {Jang}, {Jawor}, {Janssen}, {Jessner}, {Joshi}, {Kareem}, {Karuppusamy},
  {Keane}, {Keith}, {Kharbanda}, {Khizriev}, {Kikunaga}, {Kolhe}, {Kramer},
  {Krishnakumar}, {Lackeos}, {Lee}, {Liu}, {Liu}, {Lyne}, {McKee}, {Maan},
  {Main}, {Mickaliger}, {Middleton}, {Neronov}, {Nitu}, {Nobleson}, {Paladi},
  {Parthasarathy}, {Perera}, {Perrodin}, {Petiteau}, {Porayko}, {Possenti},
  {Prabu}, {Postnov}, {Quelquejay Leclere}, {Rana}, {Roper Pol}, {Samajdar},
  {Sanidas}, {Semikoz}, {Sesana}, {Shaifullah}, {Singha}, {Smarra}, {Speri},
  {Spiewak}, {Srivastava}, {Stappers}, {Steer}, {Surnis}, {Susarla},
  {Susobhanan}, {Takahashi}, {Tarafdar}, {Theureau}, {Tiburzi}, {Truant}, {van
  der Wateren}, {Valtolina}, {Vecchio}, {Venkatraman Krishnan}, {Verbiest},
  {Wang}, {Wang}, \& {Wu}}]{2023arXiv230616227A}
{Antoniadis}, J., {Arumugam}, P., {Arumugam}, S., {et~al.} 2023{\natexlab{a}},
  arXiv e-prints, arXiv:2306.16227

\bibitem[{{Antoniadis} {et~al.}(2023{\natexlab{b}}){Antoniadis}, {Arumugam},
  {Arumugam}, {Babak}, {Bagchi}, {Bak Nielsen}, {Bassa}, {Bathula},
  {Berthereau}, {Bonetti}, {Bortolas}, {Brook}, {Burgay}, {Caballero},
  {Chalumeau}, {Champion}, {Chanlaridis}, {Chen}, {Cognard}, {Dandapat}, {Deb},
  {Desai}, {Desvignes}, {Dhanda-Batra}, {Dwivedi}, {Falxa}, {Ferdman},
  {Franchini}, {Gair}, {Goncharov}, {Gopakumar}, {Graikou}, {Grie{\ss}meier},
  {Guillemot}, {Guo}, {Gupta}, {Hisano}, {Hu}, {Iraci}, {Izquierdo-Villalba},
  {Jang}, {Jawor}, {Janssen}, {Jessner}, {Joshi}, {Kareem}, {Karuppusamy},
  {Keane}, {Keith}, {Kharbanda}, {Kikunaga}, {Kolhe}, {Kramer}, {Krishnakumar},
  {Lackeos}, {Lee}, {Liu}, {Liu}, {Lyne}, {McKee}, {Maan}, {Main},
  {Mickaliger}, {Nitu}, {Nobleson}, {Paladi}, {Parthasarathy}, {Perera},
  {Perrodin}, {Petiteau}, {Porayko}, {Possenti}, {Prabu}, {Quelquejay Leclere},
  {Rana}, {Samajdar}, {Sanidas}, {Sesana}, {Shaifullah}, {Singha}, {Speri},
  {Spiewak}, {Srivastava}, {Stappers}, {Surnis}, {Susarla}, {Susobhanan},
  {Takahashi}, {Tarafdar}, {Theureau}, {Tiburzi}, {van der Wateren}, {Vecchio},
  {Venkatraman Krishnan}, {Verbiest}, {Wang}, {Wang}, \&
  {Wu}}]{2023arXiv230616214A}
{Antoniadis}, J., {Arumugam}, P., {Arumugam}, S., {et~al.} 2023{\natexlab{b}},
  arXiv e-prints, arXiv:2306.16214

\bibitem[{{Antoniadis} {et~al.}(2023{\natexlab{c}}){Antoniadis}, {Arumugam},
  {Arumugam}, {Babak}, {Bagchi}, {Bak Nielsen}, {Bassa}, {Bathula},
  {Berthereau}, {Bonetti}, {Bortolas}, {Brook}, {Burgay}, {Caballero},
  {Chalumeau}, {Champion}, {Chanlaridis}, {Chen}, {Cognard}, {Dandapat}, {Deb},
  {Desai}, {Desvignes}, {Dhanda-Batra}, {Dwivedi}, {Falxa}, {Ferdman},
  {Franchini}, {Gair}, {Goncharov}, {Gopakumar}, {Graikou}, {Grie{\ss}meier},
  {Guillemot}, {Guo}, {Gupta}, {Hisano}, {Hu}, {Iraci}, {Izquierdo-Villalba},
  {Jang}, {Jawor}, {Janssen}, {Jessner}, {Joshi}, {Kareem}, {Karuppusamy},
  {Keane}, {Keith}, {Kharbanda}, {Kikunaga}, {Kolhe}, {Kramer}, {Krishnakumar},
  {Lackeos}, {Lee}, {Liu}, {Liu}, {Lyne}, {McKee}, {Maan}, {Main},
  {Mickaliger}, {Ni{\c{t}}u}, {Nobleson}, {Paladi}, {Parthasarathy}, {Perera},
  {Perrodin}, {Petiteau}, {Porayko}, {Possenti}, {Prabu}, {Quelquejay Leclere},
  {Rana}, {Samajdar}, {Sanidas}, {Sesana}, {Shaifullah}, {Singha}, {Speri},
  {Spiewak}, {Srivastava}, {Stappers}, {Surnis}, {Susarla}, {Susobhanan},
  {Takahashi}, {Tarafdar}, {Theureau}, {Tiburzi}, {van der Wateren}, {Vecchio},
  {Venkatraman Krishnan}, {Verbiest}, {Wang}, {Wang}, \&
  {Wu}}]{2023arXiv230616225A}
{Antoniadis}, J., {Arumugam}, P., {Arumugam}, S., {et~al.} 2023{\natexlab{c}},
  arXiv e-prints, arXiv:2306.16225

\bibitem[{{Antoniadis} {et~al.}(2023{\natexlab{d}}){Antoniadis}, {Arumugam},
  {Arumugam}, {Babak}, {Bagchi}, {Bak Nielsen}, {Bassa}, {Bathula},
  {Berthereau}, {Bonetti}, {Bortolas}, {Brook}, {Burgay}, {Caballero},
  {Chalumeau}, {Champion}, {Chanlaridis}, {Chen}, {Cognard}, {Dandapat}, {Deb},
  {Desai}, {Desvignes}, {Dhanda-Batra}, {Dwivedi}, {Falxa}, {Ferranti},
  {Ferdman}, {Franchini}, {Gair}, {Goncharov}, {Gopakumar}, {Graikou},
  {Grie{\ss}meier}, {Guillemot}, {Guo}, {Gupta}, {Hisano}, {Hu}, {Iraci},
  {Izquierdo-Villalba}, {Jang}, {Jawor}, {Janssen}, {Jessner}, {Joshi},
  {Kareem}, {Karuppusamy}, {Keane}, {Keith}, {Kharbanda}, {Kikunaga}, {Kolhe},
  {Kramer}, {Krishnakumar}, {Lackeos}, {Lee}, {Liu}, {Liu}, {Lyne}, {McKee},
  {Maan}, {Main}, {Manzini}, {Mickaliger}, {Nitu}, {Nobleson}, {Paladi},
  {Parthasarathy}, {Perera}, {Perrodin}, {Petiteau}, {Porayko}, {Possenti},
  {Prabu}, {Quelquejay Leclere}, {Rana}, {Samajdar}, {Sanidas}, {Sesana},
  {Shaifullah}, {Singha}, {Speri}, {Spiewak}, {Srivastava}, {Stappers},
  {Surnis}, {Susarla}, {Susobhanan}, {Takahashi}, {Tarafdar}, {Theureau},
  {Tiburzi}, {van der Wateren}, {Vecchio}, {Venkatraman Krishnan}, {Verbiest},
  {Wang}, {Wang}, \& {Wu}}]{2023arXiv230616226A}
{Antoniadis}, J., {Arumugam}, P., {Arumugam}, S., {et~al.} 2023{\natexlab{d}},
  arXiv e-prints, arXiv:2306.16226

\bibitem[{{Antoniadis} {et~al.}(2023{\natexlab{e}}){Antoniadis}, {Babak}, {Bak
  Nielsen}, {Bassa}, {Berthereau}, {Bonetti}, {Bortolas}, {Brook}, {Burgay},
  {Caballero}, {Chalumeau}, {Champion}, {Chanlaridis}, {Chen}, {Cognard},
  {Desvignes}, {Falxa}, {Ferdman}, {Franchini}, {Gair}, {Goncharov}, {Graikou},
  {Grie{\ss}meier}, {Guillemot}, {Guo}, {Hu}, {Iraci}, {Izquierdo-Villalba},
  {Jang}, {Jawor}, {Janssen}, {Jessner}, {Karuppusamy}, {Keane}, {Keith},
  {Kramer}, {Krishnakumar}, {Lackeos}, {Lee}, {Liu}, {Liu}, {Lyne}, {McKee},
  {Main}, {Mickaliger}, {Nitu}, {Parthasarathy}, {Perera}, {Perrodin},
  {Petiteau}, {Porayko}, {Possenti}, {Samajdar}, {Sanidas}, {Sesana},
  {Shaifullah}, {Speri}, {Spiewak}, {Stappers}, {Susarla}, {Theureau},
  {Tiburzi}, {van der Wateren}, {Vecchio}, {Venkatraman Krishnan}, {Verbiest},
  {Wang}, {Wang}, \& {Wu}}]{2023arXiv230616224A}
{Antoniadis}, J., {Babak}, S., {Bak Nielsen}, A.~S., {et~al.}
  2023{\natexlab{e}}, arXiv e-prints, arXiv:2306.16224

\bibitem[{{Armitage} \& {Natarajan}(2002)}]{ArmitageNatarajan2002}
{Armitage}, P.~J. \& {Natarajan}, P. 2002, \apjl, 567, L9

\bibitem[{{Artymowicz} {et~al.}(1991){Artymowicz}, {Clarke}, {Lubow}, \&
  {Pringle}}]{Artymowicz1991}
{Artymowicz}, P., {Clarke}, C.~J., {Lubow}, S.~H., \& {Pringle}, J.~E. 1991,
  \apjl, 370, L35

\bibitem[{{Artymowicz} \& {Lubow}(1994)}]{artymowicz1994}
{Artymowicz}, P. \& {Lubow}, S.~H. 1994, \apj, 421, 651

\bibitem[{{Artymowicz} \& {Lubow}(1996)}]{artymowicz1996}
{Artymowicz}, P. \& {Lubow}, S.~H. 1996, \apjl, 467, L77

\bibitem[{{Balbus} \& {Hawley}(1991)}]{balbushawley1991}
{Balbus}, S.~A. \& {Hawley}, J.~F. 1991, \apj, 376, 214

\bibitem[{Barrows {et~al.}(2011)Barrows, Stern, Madsen, Harrison, Assef,
  Comerford, Cushing, Fassnacht, Gonzalez, Griffith, Hickox, Kirkpatrick, \&
  Lagattuta}]{Barrows2011}
Barrows, R.~S., Stern, D., Madsen, K., {et~al.} 2011, The Astrophysical
  Journal, 744, 7

\bibitem[{{Bate} {et~al.}(1995){Bate}, {Bonnell}, \& {Price}}]{bate1995}
{Bate}, M.~R., {Bonnell}, I.~A., \& {Price}, N.~M. 1995, \mnras, 277, 362

\bibitem[{{Begelman} {et~al.}(1980){Begelman}, {Blandford}, \&
  {Rees}}]{Begelman1980}
{Begelman}, M.~C., {Blandford}, R.~D., \& {Rees}, M.~J. 1980, \nat, 287, 307

\bibitem[{Bogdanovi{\'{c}} {et~al.}(2009)Bogdanovi{\'{c}}, Eracleous, \&
  Sigurdsson}]{Bogdanovic2009}
Bogdanovi{\'{c}}, T., Eracleous, M., \& Sigurdsson, S. 2009, New Astronomy
  Reviews, 53, 113

\bibitem[{{Bogdanovi{\'c}} {et~al.}(2022){Bogdanovi{\'c}}, {Miller}, \&
  {Blecha}}]{Bogdanovic2022}
{Bogdanovi{\'c}}, T., {Miller}, M.~C., \& {Blecha}, L. 2022, Living Reviews in
  Relativity, 25, 3

\bibitem[{Boroson \& Lauer(2009)}]{BorosonLauer2009}
Boroson, T.~A. \& Lauer, T.~R. 2009, Nature, 458, 53

\bibitem[{{Bortolas} {et~al.}(2021){Bortolas}, {Franchini}, {Bonetti}, \&
  {Sesana}}]{Bortolas2021}
{Bortolas}, E., {Franchini}, A., {Bonetti}, M., \& {Sesana}, A. 2021, \apjl,
  918, L15

\bibitem[{Bowen {et~al.}(2018)Bowen, Mewes, Campanelli, Noble, Krolik, \&
  Zilhão}]{Bowen2018}
Bowen, D.~B., Mewes, V., Campanelli, M., {et~al.} 2018, The Astrophysical
  Journal, 853, L17

\bibitem[{{Chandrasekhar}(1943)}]{Chandrasekhar1943}
{Chandrasekhar}, S. 1943, \apj, 97, 255

\bibitem[{{Charisi} {et~al.}(2016){Charisi}, {Bartos}, {Haiman},
  {Price-Whelan}, {Graham}, {Bellm}, {Laher}, \& {M{\'a}rka}}]{Charisi2016}
{Charisi}, M., {Bartos}, I., {Haiman}, Z., {et~al.} 2016, \mnras, 463, 2145

\bibitem[{{Chen} {et~al.}(2020){Chen}, {Liu}, {Liao}, {Holgado}, {Guo},
  {Gruendl}, {Morganson}, {Shen}, {Zhang}, {Abbott}, {Aguena}, {Allam},
  {Avila}, {Bertin}, {Bhargava}, {Brooks}, {Burke}, {Carnero Rosell},
  {Carollo}, {Carrasco Kind}, {Carretero}, {Costanzi}, {da Costa}, {Davis}, {De
  Vicente}, {Desai}, {Diehl}, {Doel}, {Everett}, {Flaugher}, {Friedel},
  {Frieman}, {Garc{\'\i}a-Bellido}, {Gaztanaga}, {Glazebrook}, {Gruen},
  {Gutierrez}, {Hinton}, {Hollowood}, {James}, {Kim}, {Kuehn}, {Kuropatkin},
  {Lewis}, {Lidman}, {Lima}, {Maia}, {March}, {Marshall}, {Menanteau},
  {Miquel}, {Palmese}, {Paz-Chinch{\'o}n}, {Plazas}, {Sanchez}, {Schubnell},
  {Serrano}, {Sevilla-Noarbe}, {Smith}, {Suchyta}, {Swanson}, {Tarle},
  {Tucker}, {Norbert Varga}, \& {Walker}}]{2020MNRAS.499.2245C}
{Chen}, Y.-C., {Liu}, X., {Liao}, W.-T., {et~al.} 2020, \mnras, 499, 2245

\bibitem[{{Combi} {et~al.}(2022){Combi}, {Lopez Armengol}, {Campanelli},
  {Noble}, {Avara}, {Krolik}, \& {Bowen}}]{2022ApJ...928..187C}
{Combi}, L., {Lopez Armengol}, F.~G., {Campanelli}, M., {et~al.} 2022, \apj,
  928, 187

\bibitem[{{Cuadra} {et~al.}(2009){Cuadra}, {Armitage}, {Alexander}, \&
  {Begelman}}]{Cuadra2009}
{Cuadra}, J., {Armitage}, P.~J., {Alexander}, R.~D., \& {Begelman}, M.~C. 2009,
  \mnras, 393, 1423

\bibitem[{{d'Ascoli} {et~al.}(2018){d'Ascoli}, {Noble}, {Bowen}, {Campanelli},
  {Krolik}, \& {Mewes}}]{Ascoli2018}
{d'Ascoli}, S., {Noble}, S.~C., {Bowen}, D.~B., {et~al.} 2018, \apj, 865, 140

\bibitem[{Decarli {et~al.}(2010)Decarli, Dotti, Montuori, Liimets, \&
  Ederoclite}]{Decarli2010}
Decarli, R., Dotti, M., Montuori, C., Liimets, T., \& Ederoclite, A. 2010, The
  Astrophysical Journal, 720, L93

\bibitem[{{D'Orazio} {et~al.}(2013){D'Orazio}, {Haiman}, \&
  {MacFadyen}}]{dorazio2013}
{D'Orazio}, D.~J., {Haiman}, Z., \& {MacFadyen}, A. 2013, \mnras, 436, 2997

\bibitem[{Dotti {et~al.}(2009)Dotti, Montuori, Decarli, Volonteri, Colpi, \&
  Haardt}]{Dotti2009}
Dotti, M., Montuori, C., Decarli, R., {et~al.} 2009, Monthly Notices of the
  Royal Astronomical Society: Letters, 398, L73

\bibitem[{{Dotti} {et~al.}(2012){Dotti}, {Sesana}, \& {Decarli}}]{Dotti2012}
{Dotti}, M., {Sesana}, A., \& {Decarli}, R. 2012, Advances in Astronomy, 2012,
  940568

\bibitem[{{Duffell} {et~al.}(2020){Duffell}, {D'Orazio}, {Derdzinski},
  {Haiman}, {MacFadyen}, {Rosen}, \& {Zrake}}]{Duffell2019}
{Duffell}, P.~C., {D'Orazio}, D., {Derdzinski}, A., {et~al.} 2020, \apj, 901,
  25

\bibitem[{{Duras} {et~al.}(2020){Duras}, {Bongiorno}, {Ricci}, {Piconcelli},
  {Shankar}, {Lusso}, {Bianchi}, {Fiore}, {Maiolino}, {Marconi}, {Onori},
  {Sani}, {Schneider}, {Vignali}, \& {La Franca}}]{Duras2020}
{Duras}, F., {Bongiorno}, A., {Ricci}, F., {et~al.} 2020, \aap, 636, A73

\bibitem[{D’Orazio \& Duffell(2021)}]{DOrazio2021}
D’Orazio, D.~J. \& Duffell, P.~C. 2021, The Astrophysical Journal Letters,
  914, L21

\bibitem[{Eracleous {et~al.}(2012)Eracleous, Boroson, Halpern, \&
  Liu}]{Eracleous2012}
Eracleous, M., Boroson, T.~A., Halpern, J.~P., \& Liu, J. 2012, The
  Astrophysical Journal Supplement Series, 201, 23

\bibitem[{{Eracleous} \& {Halpern}(1994)}]{1994ApJS...90....1E}
{Eracleous}, M. \& {Halpern}, J.~P. 1994, \apjs, 90, 1

\bibitem[{{Escala} {et~al.}(2005){Escala}, {Larson}, {Coppi}, \&
  {Mardones}}]{Escala2005}
{Escala}, A., {Larson}, R.~B., {Coppi}, P.~S., \& {Mardones}, D. 2005, \apj,
  630, 152

\bibitem[{{Farris} {et~al.}(2014){Farris}, {Duffell}, {MacFadyen}, \&
  {Haiman}}]{farris2014}
{Farris}, B.~D., {Duffell}, P., {MacFadyen}, A.~I., \& {Haiman}, Z. 2014, \apj,
  783, 134

\bibitem[{{Foster} \& {Backer}(1990)}]{Foster1990}
{Foster}, R.~S. \& {Backer}, D.~C. 1990, \apj, 361, 300

\bibitem[{{Franchini} {et~al.}(2022){Franchini}, {Lupi}, \&
  {Sesana}}]{Franchini2022}
{Franchini}, A., {Lupi}, A., \& {Sesana}, A. 2022, \apjl, 929, L13

\bibitem[{{Franchini} {et~al.}(2023){Franchini}, {Lupi}, {Sesana}, \&
  {Haiman}}]{Franchini2023}
{Franchini}, A., {Lupi}, A., {Sesana}, A., \& {Haiman}, Z. 2023, \mnras, 522,
  1569

\bibitem[{{Franchini} {et~al.}(2021){Franchini}, {Sesana}, \&
  {Dotti}}]{Franchini2021}
{Franchini}, A., {Sesana}, A., \& {Dotti}, M. 2021, \mnras, 507, 1458

\bibitem[{{Frank} {et~al.}(2002){Frank}, {King}, \&
  {Raine}}]{2002apa..book.....F}
{Frank}, J., {King}, A., \& {Raine}, D.~J. 2002, {Accretion Power in
  Astrophysics: Third Edition}

\bibitem[{{Gammie}(2001)}]{Gammie2001}
{Gammie}, C.~F. 2001, \apj, 553, 174

\bibitem[{{Goldreich} \& {Tremaine}(1980)}]{goldreich1980}
{Goldreich}, P. \& {Tremaine}, S. 1980, \apj, 241, 425

\bibitem[{{Graham} {et~al.}(2015){Graham}, {Djorgovski}, {Stern}, {Glikman},
  {Drake}, {Mahabal}, {Donalek}, {Larson}, \& {Christensen}}]{Graham2015}
{Graham}, M.~J., {Djorgovski}, S.~G., {Stern}, D., {et~al.} 2015, \nat, 518, 74

\bibitem[{Hayasaki {et~al.}(2007)Hayasaki, Mineshige, \& Sudou}]{Hayasaki2007}
Hayasaki, K., Mineshige, S., \& Sudou, H. 2007, Publications of the
  Astronomical Society of Japan, 59, 427

\bibitem[{{Heath} \& {Nixon}(2020)}]{HeathNixon2020}
{Heath}, R.~M. \& {Nixon}, C.~J. 2020, \aap, 641, A64

\bibitem[{Hogg(2000)}]{Hogg2000}
Hogg, D.~W. 2000, Distance measures in cosmology

\bibitem[{Hopkins(2015)}]{Hopkins2015}
Hopkins, P.~F. 2015, Monthly Notices of the Royal Astronomical Society, 450,
  53–110

\bibitem[{{Kelley} {et~al.}(2019){Kelley}, {Haiman}, {Sesana}, \&
  {Hernquist}}]{Kelly2019}
{Kelley}, L.~Z., {Haiman}, Z., {Sesana}, A., \& {Hernquist}, L. 2019, \mnras,
  485, 1579

\bibitem[{{Khan} {et~al.}(2011){Khan}, {Just}, \& {Merritt}}]{Khan2011}
{Khan}, F.~M., {Just}, A., \& {Merritt}, D. 2011, \apj, 732, 89

\bibitem[{{Kormendy} \& {Ho}(2013)}]{Kormendy2013}
{Kormendy}, J. \& {Ho}, L.~C. 2013, \araa, 51, 511

\bibitem[{Krolik(2010)}]{Krolik2010}
Krolik, J.~H. 2010, The Astrophysical Journal, 709, 774–779

\bibitem[{{Liu} {et~al.}(2019){Liu}, {Gezari}, {Ayers}, {Burgett}, {Chambers},
  {Hodapp}, {Huber}, {Kudritzki}, {Metcalfe}, {Tonry}, {Wainscoat}, \&
  {Waters}}]{Liu2019}
{Liu}, T., {Gezari}, S., {Ayers}, M., {et~al.} 2019, \apj, 884, 36

\bibitem[{{Lodato}(2007)}]{lodato2007sg}
{Lodato}, G. 2007, Nuovo Cimento Rivista Serie, 30 [\eprint[arXiv]{0801.3848}]

\bibitem[{{Lodato} {et~al.}(2009){Lodato}, {Nayakshin}, {King}, \&
  {Pringle}}]{Lodato2009}
{Lodato}, G., {Nayakshin}, S., {King}, A.~R., \& {Pringle}, J.~E. 2009, \mnras,
  398, 1392

\bibitem[{{Lodato} \& {Rice}(2004)}]{lodatorice2004}
{Lodato}, G. \& {Rice}, W.~K.~M. 2004, \mnras, 351, 630

\bibitem[{{Lodato} \& {Rice}(2005)}]{Lodatorice2005}
{Lodato}, G. \& {Rice}, W.~K.~M. 2005, \mnras, 358, 1489

\bibitem[{{LSST Science Collaboration} {et~al.}(2009){LSST Science
  Collaboration}, {Abell}, {Allison}, {Anderson}, {Andrew}, {Angel}, {Armus},
  {Arnett}, {Asztalos}, {Axelrod}, {Bailey}, {Ballantyne}, {Bankert},
  {Barkhouse}, {Barr}, {Barrientos}, {Barth}, {Bartlett}, {Becker}, {Becla},
  {Beers}, {Bernstein}, {Biswas}, {Blanton}, {Bloom}, {Bochanski}, {Boeshaar},
  {Borne}, {Bradac}, {Brandt}, {Bridge}, {Brown}, {Brunner}, {Bullock},
  {Burgasser}, {Burge}, {Burke}, {Cargile}, {Chandrasekharan}, {Chartas},
  {Chesley}, {Chu}, {Cinabro}, {Claire}, {Claver}, {Clowe}, {Connolly}, {Cook},
  {Cooke}, {Cooray}, {Covey}, {Culliton}, {de Jong}, {de Vries}, {Debattista},
  {Delgado}, {Dell'Antonio}, {Dhital}, {Di Stefano}, {Dickinson}, {Dilday},
  {Djorgovski}, {Dobler}, {Donalek}, {Dubois-Felsmann}, {Durech},
  {Eliasdottir}, {Eracleous}, {Eyer}, {Falco}, {Fan}, {Fassnacht}, {Ferguson},
  {Fernandez}, {Fields}, {Finkbeiner}, {Figueroa}, {Fox}, {Francke}, {Frank},
  {Frieman}, {Fromenteau}, {Furqan}, {Galaz}, {Gal-Yam}, {Garnavich},
  {Gawiser}, {Geary}, {Gee}, {Gibson}, {Gilmore}, {Grace}, {Green}, {Gressler},
  {Grillmair}, {Habib}, {Haggerty}, {Hamuy}, {Harris}, {Hawley}, {Heavens},
  {Hebb}, {Henry}, {Hileman}, {Hilton}, {Hoadley}, {Holberg}, {Holman},
  {Howell}, {Infante}, {Ivezic}, {Jacoby}, {Jain}, {R}, {Jedicke}, {Jee},
  {Garrett Jernigan}, {Jha}, {Johnston}, {Jones}, {Juric}, {Kaasalainen},
  {Styliani}, {Kafka}, {Kahn}, {Kaib}, {Kalirai}, {Kantor}, {Kasliwal},
  {Keeton}, {Kessler}, {Knezevic}, {Kowalski}, {Krabbendam}, {Krughoff},
  {Kulkarni}, {Kuhlman}, {Lacy}, {Lepine}, {Liang}, {Lien}, {Lira}, {Long},
  {Lorenz}, {Lotz}, {Lupton}, {Lutz}, {Macri}, {Mahabal}, {Mandelbaum},
  {Marshall}, {May}, {McGehee}, {Meadows}, {Meert}, {Milani}, {Miller},
  {Miller}, {Mills}, {Minniti}, {Monet}, {Mukadam}, {Nakar}, {Neill}, {Newman},
  {Nikolaev}, {Nordby}, {O'Connor}, {Oguri}, {Oliver}, {Olivier}, {Olsen},
  {Olsen}, {Olszewski}, {Oluseyi}, {Padilla}, {Parker}, {Pepper}, {Peterson},
  {Petry}, {Pinto}, {Pizagno}, {Popescu}, {Prsa}, {Radcka}, {Raddick},
  {Rasmussen}, {Rau}, {Rho}, {Rhoads}, {Richards}, {Ridgway}, {Robertson},
  {Roskar}, {Saha}, {Sarajedini}, {Scannapieco}, {Schalk}, {Schindler},
  {Schmidt}, {Schmidt}, {Schneider}, {Schumacher}, {Scranton}, {Sebag},
  {Seppala}, {Shemmer}, {Simon}, {Sivertz}, {Smith}, {Allyn Smith}, {Smith},
  {Spitz}, {Stanford}, {Stassun}, {Strader}, {Strauss}, {Stubbs}, {Sweeney},
  {Szalay}, {Szkody}, {Takada}, {Thorman}, {Trilling}, {Trimble}, {Tyson}, {Van
  Berg}, {Vanden Berk}, {VanderPlas}, {Verde}, {Vrsnak}, {Walkowicz},
  {Wandelt}, {Wang}, {Wang}, {Warner}, {Wechsler}, {West}, {Wiecha},
  {Williams}, {Willman}, {Wittman}, {Wolff}, {Wood-Vasey}, {Wozniak}, {Young},
  {Zentner}, \& {Zhan}}]{LSST2009}
{LSST Science Collaboration}, {Abell}, P.~A., {Allison}, J., {et~al.} 2009,
  arXiv e-prints, arXiv:0912.0201

\bibitem[{{MacFadyen} \& {Milosavljevi{\'c}}(2008)}]{Macfadyen2008}
{MacFadyen}, A.~I. \& {Milosavljevi{\'c}}, M. 2008, \apj, 672, 83

\bibitem[{{Major Krauth} {et~al.}(2023){Major Krauth}, {Davelaar}, {Haiman},
  {Westernacher-Schneider}, {Zrake}, \& {MacFadyen}}]{MajorKrauth2023}
{Major Krauth}, L., {Davelaar}, J., {Haiman}, Z., {et~al.} 2023, arXiv
  e-prints, arXiv:2304.02575

\bibitem[{{Mayer} {et~al.}(2007){Mayer}, {Kazantzidis}, {Madau}, {Colpi},
  {Quinn}, \& {Wadsley}}]{Mayer2007b}
{Mayer}, L., {Kazantzidis}, S., {Madau}, P., {et~al.} 2007, Science, 316, 1874

\bibitem[{{Moody} {et~al.}(2019){Moody}, {Shi}, \& {Stone}}]{Moody2019}
{Moody}, M. S.~L., {Shi}, J.-M., \& {Stone}, J.~M. 2019, \apj, 875, 66

\bibitem[{{Mu{\~n}oz} {et~al.}(2020){Mu{\~n}oz}, {Lai}, {Kratter}, \& {Mirand
  a}}]{Munoz2020}
{Mu{\~n}oz}, D.~J., {Lai}, D., {Kratter}, K., \& {Mirand a}, R. 2020, \apj,
  889, 114

\bibitem[{{Mu{\~n}oz} {et~al.}(2019){Mu{\~n}oz}, {Miranda}, \&
  {Lai}}]{Munoz2019}
{Mu{\~n}oz}, D.~J., {Miranda}, R., \& {Lai}, D. 2019, \apj, 871, 84

\bibitem[{Noble {et~al.}(2012)Noble, Mundim, Nakano, Krolik, Campanelli,
  Zlochower, \& Yunes}]{Noble2012}
Noble, S.~C., Mundim, B.~C., Nakano, H., {et~al.} 2012, The Astrophysical
  Journal, 755, 51

\bibitem[{{Pe{\~n}il} {et~al.}(2022){Pe{\~n}il}, {Ajello}, {Buson},
  {Dom{\'\i}nguez}, {Westernacher-Schneider}, \& {Zrake}}]{2022arXiv221101894P}
{Pe{\~n}il}, P., {Ajello}, M., {Buson}, S., {et~al.} 2022, arXiv e-prints,
  arXiv:2211.01894

\bibitem[{{Preto} {et~al.}(2011){Preto}, {Berentzen}, {Berczik}, \&
  {Spurzem}}]{2011ApJ...732L..26P}
{Preto}, M., {Berentzen}, I., {Berczik}, P., \& {Spurzem}, R. 2011, \apjl, 732,
  L26

\bibitem[{{Price}(2007)}]{Price2007}
{Price}, D.~J. 2007, \pasa, 24, 159

\bibitem[{{Price} {et~al.}(2017){Price}, {Wurster}, {Nixon}, {Tricco},
  {Toupin}, {Pettitt}, {Chan}, {Laibe}, {Glover}, {Dobbs}, {Nealon}, {Liptai},
  {Worpel}, {Bonnerot}, {Dipierro}, {Ragusa}, {Federrath}, {Iaconi},
  {Reichardt}, {Forgan}, {Hutchison}, {Constantino}, {Ayliffe}, {Mentiplay},
  {Hirsh}, \& {Lodato}}]{Price2017}
{Price}, D.~J., {Wurster}, J., {Nixon}, C., {et~al.} 2017, {PHANTOM: Smoothed
  particle hydrodynamics and magnetohydrodynamics code}, Astrophysics Source
  Code Library

\bibitem[{{Quinlan}(1996)}]{Quinlan1996}
{Quinlan}, G.~D. 1996, \na, 1, 35

\bibitem[{{Reardon} {et~al.}(2023){Reardon}, {Zic}, {Shannon}, {Hobbs},
  {Bailes}, {Di Marco}, {Kapur}, {Rogers}, {Thrane}, {Askew}, {Bhat},
  {Cameron}, {Cury{\l}o}, {Coles}, {Dai}, {Goncharov}, {Kerr}, {Kulkarni},
  {Levin}, {Lower}, {Manchester}, {Mandow}, {Miles}, {Nathan}, {Os{\l}owski},
  {Russell}, {Spiewak}, {Zhang}, \& {Zhu}}]{ppta2023}
{Reardon}, D.~J., {Zic}, A., {Shannon}, R.~M., {et~al.} 2023, \apjl, 951, L6

\bibitem[{{Regan} {et~al.}(2019){Regan}, {Downes}, {Volonteri}, {Beckmann},
  {Lupi}, {Trebitsch}, \& {Dubois}}]{Regan2019}
{Regan}, J.~A., {Downes}, T.~P., {Volonteri}, M., {et~al.} 2019, \mnras, 486,
  3892

\bibitem[{{Roedig} {et~al.}(2011){Roedig}, {Dotti}, {Sesana}, {Cuadra}, \&
  {Colpi}}]{roedig2011}
{Roedig}, C., {Dotti}, M., {Sesana}, A., {Cuadra}, J., \& {Colpi}, M. 2011,
  \mnras, 415, 3033

\bibitem[{{Roedig} \& {Sesana}(2014)}]{roedig2014}
{Roedig}, C. \& {Sesana}, A. 2014, \mnras, 439, 3476

\bibitem[{{Roedig} {et~al.}(2012){Roedig}, {Sesana}, {Dotti}, {Cuadra},
  {Amaro-Seoane}, \& {Haardt}}]{roedig2012}
{Roedig}, C., {Sesana}, A., {Dotti}, M., {et~al.} 2012, \aap, 545, A127

\bibitem[{{Sandrinelli} {et~al.}(2018){Sandrinelli}, {Covino}, {Treves},
  {Holgado}, {Sesana}, {Lindfors}, \& {Ramazani}}]{2018A&A...615A.118S}
{Sandrinelli}, A., {Covino}, S., {Treves}, A., {et~al.} 2018, \aap, 615, A118

\bibitem[{{Sesana} {et~al.}(2007){Sesana}, {Haardt}, \& {Madau}}]{Sesana2007}
{Sesana}, A., {Haardt}, F., \& {Madau}, P. 2007, \apj, 660, 546

\bibitem[{{Sesana} {et~al.}(2008){Sesana}, {Vecchio}, \&
  {Colacino}}]{Sesana2008}
{Sesana}, A., {Vecchio}, A., \& {Colacino}, C.~N. 2008, \mnras, 390, 192

\bibitem[{{Shakura} \& {Sunyaev}(1973)}]{ShakuraSunyaev1973}
{Shakura}, N.~I. \& {Sunyaev}, R.~A. 1973, \aap, 24, 337

\bibitem[{{Shi} {et~al.}(2012){Shi}, {Krolik}, {Lubow}, \& {Hawley}}]{Shi2012}
{Shi}, J.-M., {Krolik}, J.~H., {Lubow}, S.~H., \& {Hawley}, J.~F. 2012, \apj,
  749, 118

\bibitem[{{Siwek} {et~al.}(2023){Siwek}, {Weinberger}, \&
  {Hernquist}}]{Siwek2023}
{Siwek}, M., {Weinberger}, R., \& {Hernquist}, L. 2023, \mnras, 522, 2707

\bibitem[{{Smarra} {et~al.}(2023){Smarra}, {Goncharov}, {Barausse},
  {Antoniadis}, {Babak}, {Bak Nielsen}, {Bassa}, {Berthereau}, {Bonetti},
  {Bortolas}, {Brook}, {Burgay}, {Caballero}, {Chalumeau}, {Champion},
  {Chanlaridis}, {Chen}, {Cognard}, {Desvignes}, {Falxa}, {Ferdman},
  {Franchini}, {Gair}, {Graikou}, {Grie}, {Guillemot}, {Guo}, {Hu}, {Iraci},
  {Izquierdo-Villalba}, {Jang}, {Jawor}, {Janssen}, {Jessner}, {Karuppusamy},
  {Keane}, {Keith}, {Kramer}, {Krishnakumar}, {Lackeos}, {Lee}, {Liu}, {Liu},
  {Lyne}, {McKee}, {Main}, {Mickaliger}, {Ni{\c{t}}u}, {Parthasarathy},
  {Perera}, {Perrodin}, {Petiteau}, {Porayko}, {Possenti}, {Quelquejay
  Leclere}, {Samajdar}, {Sanidas}, {Sesana}, {Shaifullah}, {Speri}, {Spiewak},
  {Stappers}, {Susarla}, {Theureau}, {Tiburzi}, {van der Wateren}, {Vecchio},
  {Venkatraman Krishnan}, {Wang}, {Wang}, \& {Wu}}]{Smarra2023}
{Smarra}, C., {Goncharov}, B., {Barausse}, E., {et~al.} 2023, arXiv e-prints,
  arXiv:2306.16228

\bibitem[{{Tanaka} {et~al.}(2012){Tanaka}, {Perna}, \& {Haiman}}]{Tanaka2012}
{Tanaka}, T., {Perna}, R., \& {Haiman}, Z. 2012, \mnras, 425, 2974

\bibitem[{Tang \& Grindlay(2009)}]{TangGrindlay2009}
Tang, S. \& Grindlay, J. 2009, The Astrophysical Journal, 704, 1189

\bibitem[{{Tang} {et~al.}(2018){Tang}, {Haiman}, \& {MacFadyen}}]{Tang2018}
{Tang}, Y., {Haiman}, Z., \& {MacFadyen}, A. 2018, \mnras, 476, 2249

\bibitem[{{Tiede} {et~al.}(2020){Tiede}, {Zrake}, {MacFadyen}, \&
  {Haiman}}]{Tiede2020}
{Tiede}, C., {Zrake}, J., {MacFadyen}, A., \& {Haiman}, Z. 2020, \apj, 900, 43

\bibitem[{{Toomre}(1964)}]{Toomre1964}
{Toomre}, A. 1964, \apj, 139, 1217

\bibitem[{{Tsai} {et~al.}(2013){Tsai}, {Jarrett}, {Stern}, {Emonts}, {Barrows},
  {Assef}, {Norris}, {Eisenhardt}, {Lonsdale}, {Blain}, {Benford}, {Wu},
  {Stalder}, {Stubbs}, {High}, {Li}, \& {Kong}}]{Tsai2013}
{Tsai}, C.-W., {Jarrett}, T.~H., {Stern}, D., {et~al.} 2013, \apj, 779, 41

\bibitem[{{Tsalmantza} {et~al.}(2011){Tsalmantza}, {Decarli}, {Dotti}, \&
  {Hogg}}]{Tsalmantza2011}
{Tsalmantza}, P., {Decarli}, R., {Dotti}, M., \& {Hogg}, D.~W. 2011, \apj, 738,
  20

\bibitem[{{Vaughan} {et~al.}(2016){Vaughan}, {Uttley}, {Markowitz},
  {Huppenkothen}, {Middleton}, {Alston}, {Scargle}, \&
  {Farr}}]{2016MNRAS.461.3145V}
{Vaughan}, S., {Uttley}, P., {Markowitz}, A.~G., {et~al.} 2016, \mnras, 461,
  3145

\bibitem[{Westernacher-Schneider {et~al.}(2022)Westernacher-Schneider, Zrake,
  MacFadyen, \& Haiman}]{Westernacher2022}
Westernacher-Schneider, J.~R., Zrake, J., MacFadyen, A., \& Haiman, Z. 2022,
  Physical Review D, 106

\bibitem[{Westernacher-Schneider {et~al.}(2023)Westernacher-Schneider, Zrake,
  MacFadyen, \& Haiman}]{Westernacher2023}
Westernacher-Schneider, J.~R., Zrake, J., MacFadyen, A., \& Haiman, Z. 2023,
  Characteristic signatures of accreting binary black holes produced by
  eccentric minidisks

\bibitem[{{Xu} {et~al.}(2023){Xu}, {Chen}, {Guo}, {Jiang}, {Wang}, {Xu}, {Xue},
  {Nicolas Caballero}, {Yuan}, {Xu}, {Wang}, {Hao}, {Luo}, {Lee}, {Han},
  {Jiang}, {Shen}, {Wang}, {Wang}, {Xu}, {Wu}, {Manchester}, {Qian}, {Guan},
  {Huang}, {Sun}, \& {Zhu}}]{cpta2023}
{Xu}, H., {Chen}, S., {Guo}, Y., {et~al.} 2023, Research in Astronomy and
  Astrophysics, 23, 075024

\bibitem[{{Zrake} {et~al.}(2021){Zrake}, {Tiede}, {MacFadyen}, \&
  {Haiman}}]{Zrake2021}
{Zrake}, J., {Tiede}, C., {MacFadyen}, A., \& {Haiman}, Z. 2021, \apjl, 909,
  L13

\end{thebibliography}
\end{document}